\documentclass[english,12pt,a4paper]{article}
\usepackage{authblk}
\usepackage[text={17.0cm,23.7cm}]{geometry}
\usepackage[T1]{fontenc}
\usepackage[utf8]{inputenc}
\usepackage{babel}
\usepackage{graphicx}
\usepackage{tabularx}
\usepackage{amsmath}
\usepackage{multirow}
\usepackage{booktabs}
\usepackage{amssymb}
\usepackage[dvipsnames]{xcolor}
\usepackage{yfonts}
\usepackage{ulem}
\usepackage{enumitem}
\usepackage{mdframed}
\usepackage{xcolor}
\usepackage[style=numeric-comp, sorting=none]{biblatex}
\usepackage{float}
\usepackage{mathrsfs}

\usepackage[colorlinks=true, allcolors=blue]{hyperref}

\begin{document}
	
	\title{\vspace{-2cm}
	{\normalsize \flushright TUM-HEP 1447/22, IPMU22-0070\\}
		\vspace{0.6cm}
	\textbf{Enhanced prospects for direct detection of inelastic dark matter from a non-galactic diffuse component}\\[8mm]}

	\author[1,2]{Gonzalo Herrera}
	\author[1]{Alejandro Ibarra}
	\author[3]{Satoshi Shirai}
	\affil[1]{\normalsize\textit{Physik-Department, Technische Universit\"at M\"unchen, James-Franck-Stra\ss{}e, 85748 Garching, Germany}}
	\affil[2]{\normalsize\textit{Max-Planck-Institut f\"ur Physik (Werner-Heisenberg-Institut), F\"ohringer Ring 6,80805 M\"unchen, Germany}}
	\affil[3]{\normalsize\textit{Kavli Institute for the Physics and Mathematics of the Universe (WPI),The University of Tokyo Institutes for Advanced Study, The University of Tokyo, Kashiwa 277-8583, Japan}}
	\date{}
	
	\maketitle
	
	\begin{abstract}
	In some scenarios, the dark matter particle predominantly scatters inelastically with the target, producing a heavier neutral particle in the final state. In this class of scenarios, the reach in parameter space of direct detection experiments is limited by the velocity of the dark matter particle, usually taken as the escape velocity from the Milky Way.  On the other hand, it has been argued that a fraction of the dark matter particles in the Solar System could  be bound to the envelope of the Local Group or to the Virgo Supercluster, and not to our Galaxy, and therefore could carry velocities larger than the escape velocity from the Milky Way. In this paper we estimate the enhancement in sensitivity of current direct detection experiments to inelastic dark matter  scatterings with nucleons or electrons due to the non-galactic diffuse components, and we discuss the implications for some well motivated models.
	\end{abstract}
	
	\section{Introduction}
	
	The existence of dark matter in galaxies, clusters of galaxies and the Universe at large scale is by now established by their gravitational effects on ordinary matter (for reviews, see {\it e.g.} \cite{Jungman:1995df,Bertone:2004pz,Bergstrom:2000pn,Feng:2010gw}). If the dark matter is constituted by new particles, it is plausible that they could interact with the ordinary matter through other interactions aside from gravity. A promising avenue to probe these putative interactions consists in the search for nuclear or electron recoils induced by dark matter particles entering a dedicated detector at the Earth~\cite{Goodman:1984dc,Bernabei:2007gr} (for reviews, see {\it e.g.} \cite{Cerdeno:2010jj,MarrodanUndagoitia:2015veg,Lewin:1995rx}). This search strategy, denominated direct detection, has seen an impressive increase in sensitivity since it was first proposed more than three decades ago. Yet, no conclusive dark matter signal has been found to date. 
	
	Assuming that the dark matter scatters elastically with the nucleus, current direct detection experiments restrict the  spin-independent interaction cross-section to be smaller than $\sim$ 1 zeptobarn in the mass range $\sim 10$ GeV - 1 TeV ~\cite{LZ:2022ufs}. These stringent constraints put pressure on several well motivated dark matter scenarios, especially those for which the dark matter particle couples at tree level with the valence quarks in models addressing the electroweak hierarchy problem~\cite{Jungman:1995df}. On the other hand, there are many other dark matter scenarios, arguably also well motivated theoretically, which are largely unconstrained by current searches.

	In this paper we will focus on scenarios where the dark matter cannot scatter elastically with a nucleus (or an electron), so that the stringent limits on the elastic scattering cross-section do not necessarily hold. This seemingly strong assumption naturally arises in some models. For instance, the elastic scattering mediated by vector current is forbidden for Majorana dark matter $\chi$, due to the Majorana nature of fermion: $\bar{\chi} \gamma^\mu \chi = 0$~\cite{Nieves:1981zt}. However, Majorana dark matter particles may leave an imprint in direct search experiments if they could scatter inelastically producing a heavier Majorana fermion $\chi'$ in the final state, since there is an off-diagonal fermion current $\bar{\chi}' \gamma^\mu \chi \ne 0$. This scenario is  approximately realized in the Minimal Supersymmetric Standard Model, when the lightest supersymmetric particle is almost a pure Higgsino state, and the other supersymmetric particles are very heavy. In this case, the elastic scattering of the Higgsino dark matter is suppressed by the large sfermion and gaugino masses, while it has a large inelastic scattering cross section by the electroweak gauge interactions~\cite{Nagata_2015}. Scenarios of inelastic dark matter have also been motivated phenomenologically, {\it e.g.} in  \cite{Hall_1998,Alves_2010,Schwetz_2011,Arkani_Hamed_2009,Chang_2010,Barello_2014,Nagata_2015,Nagata_2015_2,Emken_2022, Bell_2018, Biswas:2022cyh, Fujiwara:2022uiq}.

The kinematics of the inelastic scattering differs from the one in the elastic scenario. In order to allow the production of a heavier neutral particle in the final state, the velocity of the incoming dark matter particle must be larger than a certain threshold. Therefore, as the mass difference between the initial and final neutral particles increases, faster and faster dark matter particles are necessary in order to open kinematically the inelastic process. For dark matter particles bound to our galaxy, and which have speeds smaller than the escape velocity from the Milky Way, $v_\mathrm{esc}=544$ km/s~\cite{Smith:2006ym,Piffl:2013mla}, the inelastic scattering off a nucleus is kinematically allowed when the mass difference between the two states is $\delta m<1/2\mu v_\mathrm{esc}^2$, with $\mu$ the reduced mass of the DM-nucleus system; for the scattering off an electron, the inelastic channel is open when $\delta m<1/2\mu_{e} v_\mathrm{esc}^2-|E^{nl}|$, where $\mu_{e}$ is the reduced mass of the DM-electron system, and $|E^{nl}|$ is the binding energy of an electron in the $(n,l)$ shell of the target nucleus. In practice, experiments can only detect recoiling nuclei/ionized electrons within a given energy range, therefore the mass difference that can be probed in direct searches is smaller than this value. 

Most analyses of direct dark matter detection implicitly assume that the Milky Way is an isolated galaxy. Instead, the Milky Way is one among the various members of the Local Group, which include M31, M33 and several dwarf galaxies. It has been argued that the Local Group contains a diffuse dark matter component, which is not bound to any individual galaxy, and which is distributed roughly homogeneously over the Local Group \cite{1959ApJ...130..705K, 2008gady.book.....B, Cox:2007nt}. Notably, a non-negligible fraction of the dark matter particles in the Solar System is expected to be associated to this non-galactic diffuse component, rather than to the Milky Way halo, and could have velocities larger than the escape velocity from the Milky Way. Likewise, the Local Group is one among the many groups of galaxies embedded in the Virgo Supercluster, which could also contain a diffuse component~\cite{Makarov:2010di}. Although the fraction of dark matter particles in the Solar System associated to the Virgo Supercluster is fairly small, they have very large velocities.
As a result, the actual dark matter velocity distribution at the Solar System is qualitatively different to the one expected from the Standard Halo Model. In \cite{ Herrera_2021} it was shown that the non-galactic diffuse component enhances the prospect for detection of scenarios where the dark matter scatters elastically with nuclei or with electrons. 

In this work we will extend that analysis to scenarios of inelastic dark matter, and we will show that the mass splittings that can be probed in direct search experiments is larger than the one previously considered in the literature.

The paper is organized as follows. In section \ref{sec:2}, we present the non-galactic dark matter flux at Earth. In section \ref{sec:3}, we derive constraints on inelastic dark matter from nuclear recoil searches, and in section \ref{sec:4}, we derive constraints from electron recoil searches. Finally, in section \ref{sec:conclusions}, we present our conclusions.

\section{Dark matter flux at  Earth}
\label{sec:2}
A correct description of the dark matter flux at Earth is crucial for assessing the prospects for detection of a given dark matter model. The largest contribution to the flux is expected to arise from dark matter particles in the Milky Way halo. The local density of dark matter particles and their velocity distribution is unknown. However, it is common in the literature to adopt the Standard Halo Model (SHM), characterized by a local density $\rho^{\rm loc}_{\rm SHM}=0.3\,{\rm GeV}/{\rm cm}^{3}$ and an isotropic velocity distribution described by a Maxwell-Boltzmann distribution truncated at the escape velocity of the Milky Way \cite{Read:2014qva,Green:2011bv}.~\footnote{Simulations and various observations suggest the existence of dark matter substructures bound to the Milky Way that may induce deviations from the Maxwell-Boltzmann form at the location of the Solar System; their impact in direct detection experiments has been discussed \textit{e.g} \cite{Freese:2003na, Kelso:2016qqj, Bozorgnia:2016ogo, OHare:2018trr, OHare:2019qxc, Bozorgnia:2019mjk, Besla:2019xbx, Smith-Orlik:2023kyl, Radick:2020qip, Buch:2020xyt, Maity:2022enp}. In order to compare our results with the published results from experiments, we will simply adopt the SHM.}

In the galactic frame, the velocity distribution reads:
	\begin{align}
		\label{eq:f_MB}
		f_\mathrm{SHM} ( \vec{ v } ) = \frac { 1} { ( 2\pi \sigma _ { v } ^ { 2} ) ^ { 3/ 2} N _ { \text{esc} } } \exp \left[ - \frac {v ^ { 2} } { 2\sigma _ { v } ^ { 2} } \right]\quad \text{for } v \leq v_{\rm esc}\,,
	\end{align}
where $v=|\vec v|$, $\sigma_v \approx 156$ km/s is  the velocity dispersion \cite{Kerr:1986hz,Green:2011bv}, and $v_\mathrm{esc}=544$ km/s is the escape velocity from our Galaxy~\cite{Smith:2006ym,Piffl:2013mla}. 
	Further, $N _ { \text{esc} }$ is a normalization constant, given by:
	\begin{align}
		N _ { \text{esc} } = \operatorname{erf} \left( \frac { v _ { \text{esc} } } { \sqrt { 2} \sigma _ { v } } \right) - \sqrt { \frac { 2} { \pi } } \frac { v _ { \text{esc} } } { \sigma _ { v } } \exp \left( - \frac { v _ {\mathrm{esc}} ^ { 2} } { 2\sigma _ { v } ^ { 2} } \right)\,.
	\end{align}
	For our chosen parameters, $N_{\rm esc}\simeq 0.993$.
	The contribution to the local dark matter flux from the Milky Way halo then reads:
	\begin{align}
		{\mathscr F}_{\rm SHM}(\vec v)=  \frac{\rho_{\rm SHM}^{\rm loc}}{m_{\rm DM}} v f_{\rm SHM}(\vec v) \, .
	\end{align}
	
    It is also plausible that the dark matter flux at Earth also contains a contribution from dark matter particles not bound to the Milky Way. Astronomical observations indicate the presence of diffuse dark matter components homogeneously distributed between clusters and Superclusters of galaxies \cite{Karachentsev_2012}. Since these dark matter particles are not gravitationally bound to the Milky Way, they carry larger velocities than the escape velocity of the Milky Way. 	In this work, we consider the contribution to the dark matter flux from the Local Group and from the Virgo Supercluster. The dark matter particles from the Local Group contribute at the Solar System with a local density of $\rho_{\rm LG} \sim 10^{-2}$ GeV/cm$^{3}$, and are expected to move isotropically with a narrow velocity distribution, $\sigma_{v.{\rm LG}} \sim 20$ km/s, and  with mean velocity $v_{\rm LG}\sim 600$ km/s~\cite{Baushev:2012dm}. The contribution from the Local Group to the dark matter flux at the location of the Solar System then reads:
	\begin{align}
		{\mathscr F}_{\rm LG}(\vec v)=  \frac{\rho_{\rm LG}^{\rm loc}}{ m_{\rm DM}}  \,\frac{\delta (v - v_{\rm LG})}{4\pi v}.
	\end{align}

 Dark matter particles bound to the Virgo Supercluster give a small contribution to the local dark matter density. Observations indicate that the average density in the diffuse component of the Virgo Supercluster is close to the cosmological value $\sim 10^{-6}$ GeV/cm$^{3}$ \cite{Makarov:2010di}. However, the gravitational focusing due to the Local Group leads to an increase in the density at the location of the Sun by a factor $\sim 1+v_{\rm esc}^{2}/v_{\sigma_{\rm VS}}^{2}$, where $v_{\sigma_{\rm VS}}$ is the velocity dispersion of the dark matter particles from the Virgo Supercluster \cite{Baushev:2012dm}. This value is highly uncertain, but it is expected to be comparable to that of the observable members of the Supercluster, which ranges from $v_{\sigma_{\rm VS}} \sim 50$ km/s to $v_{\sigma_{\rm VS}} \sim 500$ km/s \cite{Karachentsev:2002wf, Makarov:2010di}.  We consider for concreteness an enhancement on the local density of dark matter particles from the Virgo Supercluster of $\sim 10$, consistent with the value of the velocity dispersion of the observable members of the Supercluster, which leads to $\rho_{\rm VG}^{\rm loc}\sim 10^{-5}$ GeV/cm$^{3}$. Current knowledge on the dark matter velocity distribution in the Virgo Supercluster is much poorer. Following \cite{Baushev:2012dm}, we assume that the dark matter particles have the typical velocities of the members of the Virgo Supercluster, corresponding to (at least) $v_{\rm VS}\sim 1000$ km/s. The contribution to the dark matter flux at the location of the Solar System from the Virgo Supercluster can then be written as:
	\begin{align}
		{\mathscr F}_{\rm VS}(\vec v)=  \frac{\rho_{\rm VS}^{\rm loc}}{m_{\rm DM}} \frac{\delta(v - v_{\rm VS})}{4\pi v}.
	\end{align}

 The total (galactic plus non-galactic) dark matter flux at the Solar System is therefore approximately given by: 
	\begin{align}
		{\mathscr F}(\vec v)={\mathscr F}_{\rm SHM}(\vec v)+{\mathscr F}_{\rm  LG}(\vec v)+{\mathscr F}_{\rm \rm VS}(\vec v).
		\label{eq:total_flux}
	\end{align}
 Following \cite{Baushev:2012dm, Herrera_2021}, we adopt values for the local density of each component  such that the total sum yields the canonical value of the local density used by  direct detection experiments $\rho^{\rm loc}=0.3$ GeV/cm$^{3}$, namely $\rho^{\rm loc}_{\rm SHM}=0.26$ GeV/cm$^{3}$ ($\sim 88 \%$), $\rho^{\rm loc}_{\rm LG}=0.037$ GeV/cm$^{3}$ ($\sim 12 \%$), and $\rho^{\rm loc}_{\rm VS}=10^{-5}$ GeV/cm$^{3}$ ($\sim 0.003 \%$).

	\section{Impact on nuclear recoils}
	\label{sec:3}
	
	The differential rate of nuclear recoils induced by inelastic up-scatterings of dark matter particles traversing a detector at the Earth is given by:
	\begin{align}
		\frac{dR}{dE_R}= \sum_i \frac{ \xi_i}{m_{A_i} } \int_{v \geq v^i_{\rm min}(E_R)} \text{d}^3 v {\mathscr F}( \vec{v} + \vec v_\odot)\, \frac{\text{d}\sigma_i}{\text{d}E_R}(v, E_R) \,.
		\label{eq:diff_scattering_rate}
	\end{align}
	Here, $\vec v$ is the dark matter velocity in the rest frame of the detector, ${\mathscr F}( \vec{v}+\vec v_\odot)$ is the dark matter flux in the detector frame, and $\vec v_\odot$ is the velocity of the Sun with respect to the Galactic frame with $|\vec v_\odot| \approx 232$ km/s~\cite{McCabe:2013kea}.  For the inelastic scattering with mass splitting between two dark matter states, $\delta_{\rm DM}$, the minimum velocity necessary to induce a recoil with energy $E_R$ of the nucleus $i$ with mass $m_{A_i}$ and mass fraction $\xi_{i}$ in the detector reads
	\begin{align}
		v^i_{\rm min}(E_R) = \frac{1}{\sqrt{2 E_{R} m_{A_{i}}}}\Big(\frac{E_R m_{A_i}}{\mu_{A_i}}+\delta_{\rm DM}\Big) \, .
		\label{eq:vmin}
	\end{align} 
Further, for spin-independent interactions, the differential dark matter-nucleus cross section reads,
	\begin{align}\label{eq:sidiffcross}
		\frac{d\sigma^{\rm SI}_i}{dE_{R}}(v,E_R)=\frac{m_{A_{i}}}{2\mu_{A_{i}}^{2}v^{2}}\sigma_{0,i}^{\rm SI}F_i^{2}(E_{R})\;.
	\end{align}
	Here $m_{A_i}$ is mass of the nucleus $i$,   $\mu_{A_{i}}$ is the reduced mass of the dark matter-nucleus $i$ system and $F_i^{2}(E_{R})$ is the nuclear form-factor, for which we adopt the Helm prescription. Besides, $\sigma_{0,i}^{\rm SI}$ is the spin-independent dark matter-nucleus scattering cross section at zero momentum transfer, which depends on the details of the dark matter model and the target nucleus. From the differential rate, one can calculate the total recoil rate using:
	\begin{align}
		R=  \int_{0}^\infty \text{d}E_R \, \epsilon_i (E_R) \frac{dR}{dE_R} \, ,
		\label{eq:scattering_rate}
	\end{align}
	where $\epsilon_i(E_R)$ is the efficiency of that experiment. Finally, the total number of expected recoil events is ${\cal N}=R\cdot \mathcal{E}$, with $\mathcal{E}$ the exposure ({\it i.e.} mass multiplied by live-time).
	
In our analysis, we will consider two scenarios for the coupling of dark matter to nucleons. First, we will consider a Majorana dark matter candidate. In this case
\begin{align}
    \sigma^{\rm SI}_{0,i}= \frac{4\mu_{A_{i}}^{2}}{\pi}\Big[Z_{i} f_S^{p}+(A_{i}-Z_{i})f_S^{n}\Big]^2 \, ,
    \label{eq:sigma_scalar}
\end{align}
where $f_S^p$ and $f_S^n$ parametrize the strength of the scalar interactions to the proton and the neutron  (see {\it e.g.} \cite{Belanger:2008sj,Cerdeno:2010jj}). It is 
common to write  Eq.~(\ref{eq:sigma_scalar}) as 
\begin{align}
    \sigma^{\rm SI}_{0,i}= \frac{\mu_{A_{i}}^{2}}{\mu_p^{2}}\Big[Z_{i} +(A_{i}-Z_{i})\frac{f_S^n}{f_S^p}\Big]^2 \sigma_{\rm DM,p} \, ,
\end{align}
with $\mu_p$ the reduced mass of the DM-proton system and $\sigma_{\rm DM,p}$ an effective DM-proton interaction cross-section. Within the Majorana dark matter scenario, we will consider in particular the widely adopted benchmark case where the interaction is ``isoscalar'', {\it i.e.} when the dark matter couples with equal strength to protons and neutrons, for which
\begin{align}
    \sigma^{\rm SI}_{0,i}= \frac{\mu_{A_{i}}^{2}}{\mu_p^{2}}A_{i}^2 \sigma_{\rm DM,p} \, .
\end{align}

We will also consider a scenario where the dark matter has hypercharge $Y$, and interacts with the quarks via the exchange of a $Z$ boson. In this case, $\sigma_{0,i}^{\rm SI}$ has the same form as Eq.~(\ref{eq:sigma_scalar}), replacing the scalar couplings by the corresponding vector couplings, $f_S^{p,n}\rightarrow f_V^{p,n}$. For interactions with the $Z$ boson, $f_V^p$ and $f_V^n$ are explicitly given by:
\begin{align}
	f_V^p &= \frac{G_F \zeta Y}{2\sqrt{2}} (1-4 \sin^2\theta_W) \, ,\nonumber\\
	f_V^n &= -\frac{G_F \zeta Y}{2\sqrt{2}} \, ,
\end{align}
with $\zeta=1$ ($\zeta=2$) for fermionic (bosonic) dark matter~\cite{Goodman:1984dc,Cirelli:2005uq,Nagata_2015_2}. In this scenario, the dark matter-nucleus cross section can be related to the dark matter-proton cross-section through:
\begin{align}
\sigma^{\rm SI}_{0,i}=\frac{\mu_{A_i}^2}{\mu_p^2} \Big[ Z_i-\frac{(A_i-Z_i)}{(1-4\sin^2\theta_W)}\Big]^2\sigma_{\rm DM,p} \, ,
\end{align}
which is independent of the dark matter hypercharge and spin. 
	
\begin{figure}[t!]
  \centering   
		\includegraphics[width=0.49\textwidth]{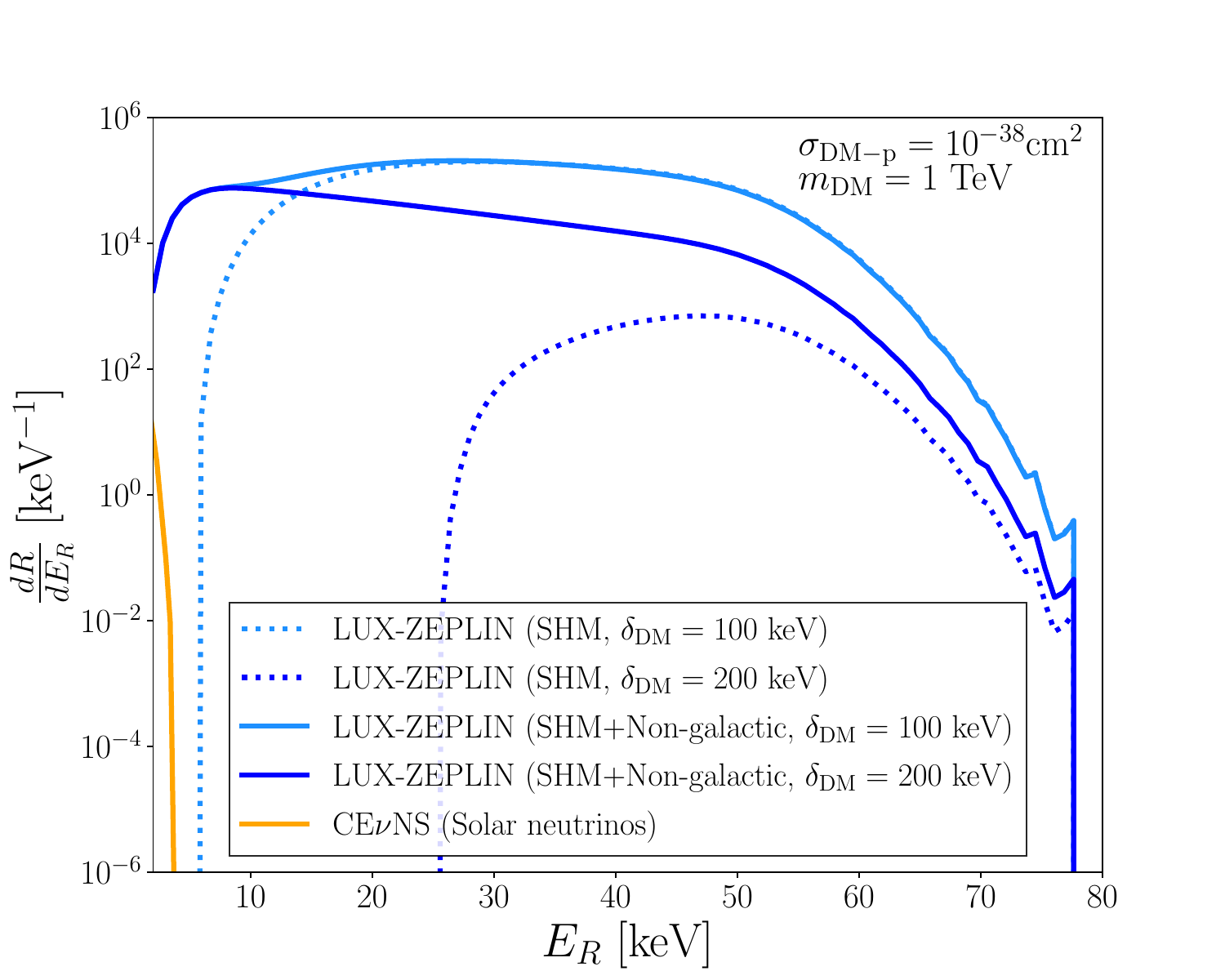}
		\caption{Differential rate for the inelastic scattering of a Majorana dark matter candidate in the ``isoscalar'' scenario with mass $m_{\rm DM}=1$ TeV, for $\delta_{\rm DM}=100$ keV (light blue) and 200 keV (dark blue), for a dark matter flux at Earth as modelled by the Standard Halo Model (dotted line) or including also the contribution from the non-galactic diffuse dark matter component (solid line). For the plots it was assumed $\sigma_{\rm DM,p}=10^{-38}\,{\rm cm}^2$.}
		\label{fig:recoil_spectrum}
	\end{figure}

To assess the impact of the non-galactic diffuse components for direct detection experiments, we plot in Figure.~\ref{fig:recoil_spectrum} the differential rate of inelastic scatterings in the LUX-ZEPLIN experiment for the ``isoscalar'' scenario, assuming $m_{\rm DM}=1$ TeV and $\sigma_{\rm DM,p}=10^{-38}\,{\rm cm}^2$, for $\delta_{\rm DM}=100$ keV (light blue) and 200 keV (dark blue), including in the flux only the contribution from dark matter bound to the Milky Way (dotted lines), as commonly assumed in the literature, and including the contribution from the non-galactic diffuse component (solid lines). The impact of the non-galactic component in the differential rate is apparent from the figure, and increases the number of events at all recoil energies, especially in the region with low $E_R$ which is not kinematically accessible to the galactic dark matter. The non-galactic dark matter, therefore, has implications not only for enhancing the sensitivity of the experiment, but also for the interpretation of a putative dark matter signal.

\begin{figure}[t!]
\centering
	\includegraphics[width=0.49\textwidth]{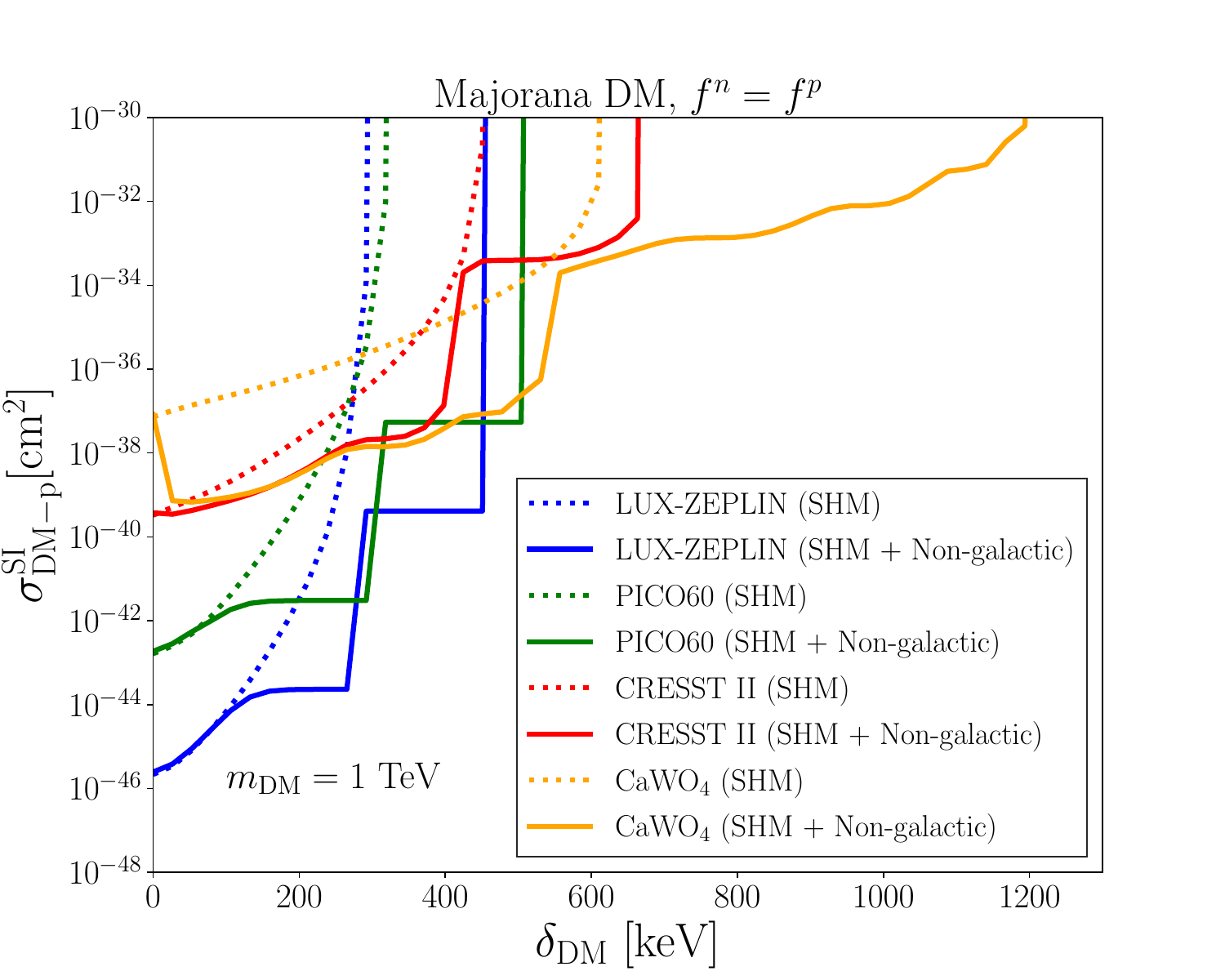} 
	\includegraphics[width=0.49\textwidth]{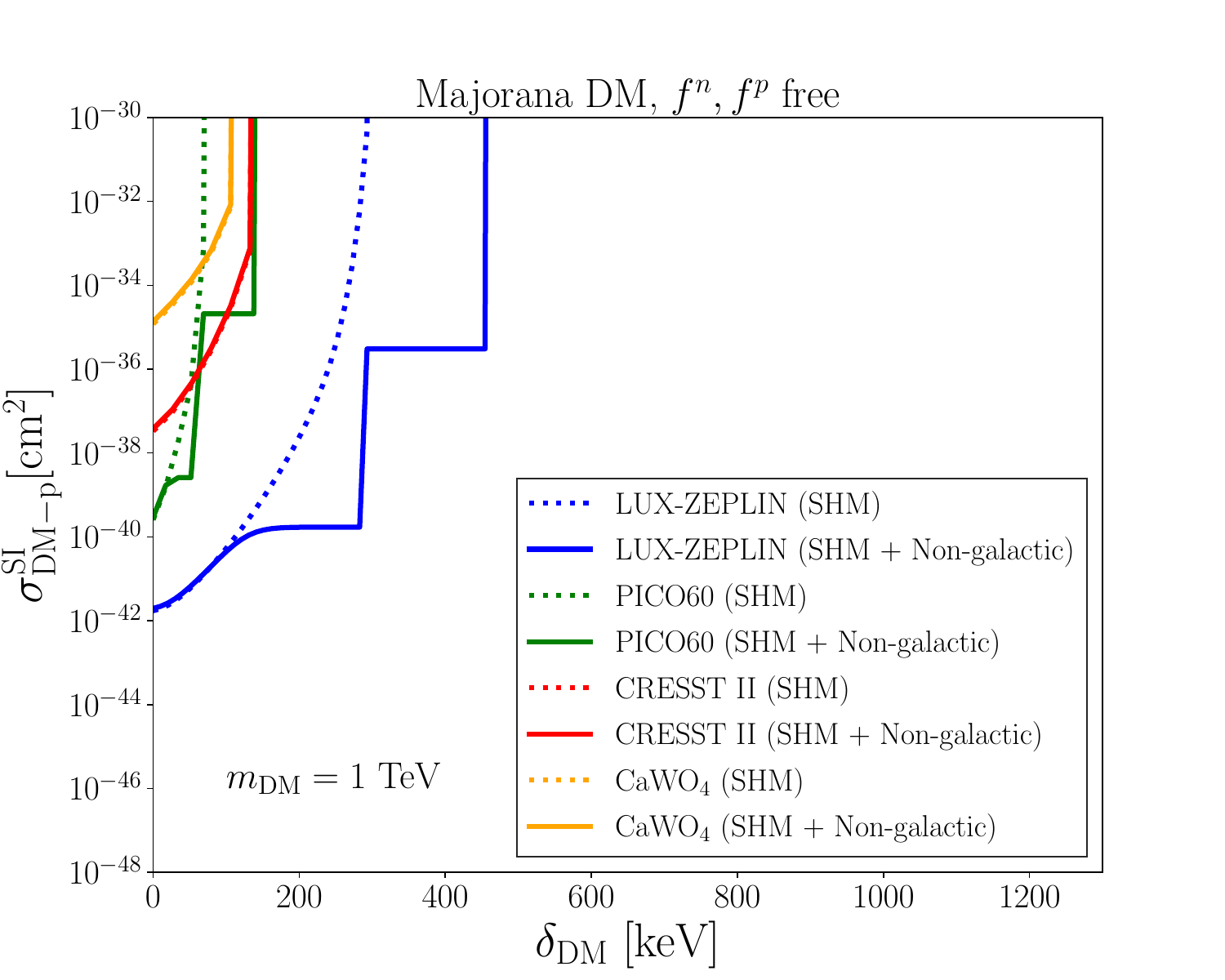}\\
	\includegraphics[width=0.49\textwidth]{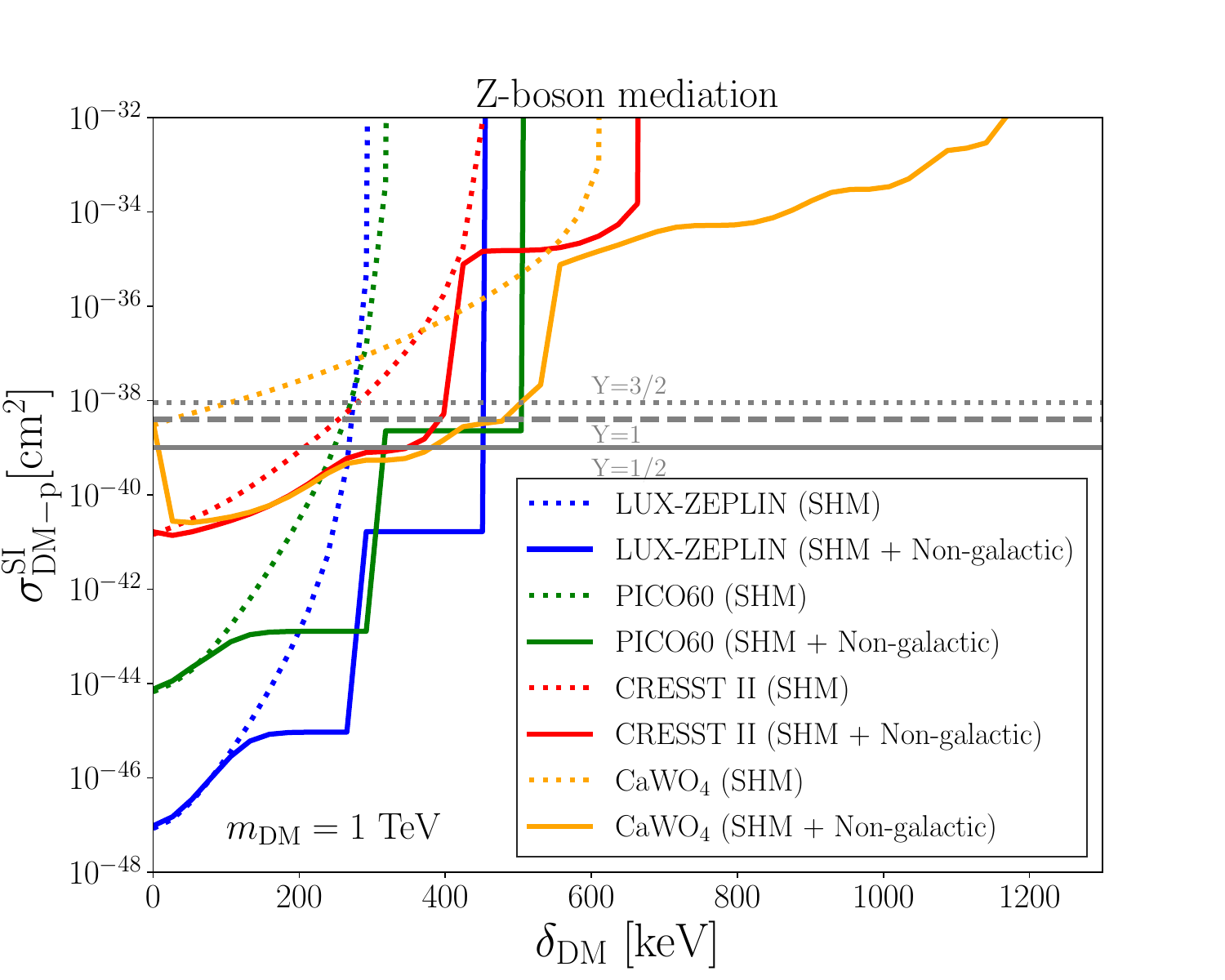}
	\caption{90$\%$ C.L upper limits on the spin-independent dark matter-proton inelastic cross section for a dark matter mass of 1 TeV as a function of the mass splitting,	from LUX-ZEPLIN (blue), PICO60 (green), CRESST-II (red and orange) and from a CaWO$_{4}$ detector radiopurity measurement (orange). We show the limits for three different scenarios: Majorana dark matter with isoscalar interactions $f^p=f^n$ (upper left plot),  arbitrary $f^p$ and $f^n$ (upper right plot), and dark matter interacting via the $Z$-boson (lower plot). In the lower plot, we also show for reference the predicted value of the cross-section with a xenon target for scenarios of fermionic dark matter with hypercharge $Y=1/2,1,3/2$.}
	\label{fig:upper_limits_spin_independent_1TeV}
\end{figure}

    Current direct search experiments have not observed a significant excess of nuclear recoils, which allows to derive upper limits on the dark matter nucleon cross section for given combinations of the dark matter mass and mass splitting between the dark matter particle and the neutral particle in the final state. In Figure \ref{fig:upper_limits_spin_independent_1TeV}, we show  upper limits on the dark matter-proton spin-independent scattering cross section versus mass splitting for $m_{\rm DM}= 1$ TeV from LUX-ZEPLIN (blue) \cite{LZ:2022ufs}, PICO60 (green) \cite{Amole_2016}, CRESST-II  (red) \cite{Angloher_2016}, and from a radiopurity measurement in a CaWO$_{4}$ crystal (orange) \cite{M_nster_2014, Song:2021yar}. The dotted lines represent the limits obtained considering the galactic dark matter (described by the SHM) as the only contribution to the dark matter flux, while the solid lines were obtained including also the contributions to the flux from the non-galactic diffuse component in the Solar System. In the upper left plot, we show the limits for a Majorana dark matter candidate in the ``isoscalar'' scenario, and in the upper right plot, the most conservative limit for the Majorana dark matter,  without making assumptions on the coupling strengths, derived following the approach of \cite{Brenner:2022qku}. Lastly, in the lower plot we show the limits for a scenario where the dark matter interacts with the nucleus via the exchange of a $Z$-boson. In the latter plot we also show the dark matter-proton scattering cross-section for scenarios of a fermionic dark matter, and $Y=1/2$ (corresponding to the well motivated scenario of the Higgsino dark matter in the limit of high scale supersymmetry~\cite{Nagata_2015}), $Y=1$ and $Y=3/2$ (which correspond to different scenarios of minimal dark matter \cite{Cirelli:2005uq}), for a xenon target. For other targets, the expected cross section for $m_{\rm DM}=1$ TeV scales as $\sim A_i/Z_i$, being indistinguishable in the Figure.

As seen in the plots, for all the scenarios the non-galactic diffuse component enhances the sensitivity of experiments to inelastic dark matter, allowing to probe larger mass splittings. For instance, for our representative dark matter mass of 1 TeV, the LUX-ZEPLIN experiment is insensitive to dark matter particles of the Milky Way scattering inelastically if the mass difference with the neutral particle in the final state is  $\delta_{\rm DM} \gtrsim 300$ keV. However, the presence of dark matter in the Solar System from the envelope of the Local Group extends the reach up to $\delta_{\rm DM}\simeq 330$ keV and allows to probe uncharted parameter space for large mass splittings. Concretely, the LUX-ZEPLIN experiment sets for the isoscalar scenario the limit $\sigma_{\rm DM-p}^{\rm SI}\lesssim 10^{-44}\,{\rm cm}^2$ for $\delta_{\rm DM} = 250$ keV, which is about three orders of magnitude stronger than the limit obtained assuming that all dark matter is bound to the Milky Way, and only a factor of 100 weaker than the limit on the elastic scattering cross-section {\it i.e.} for $\delta_{\rm DM}=0$. For the interaction mediated by the $Z$-boson the upper limit is $\sigma_{\rm DM-p}^{\rm SI}\lesssim 10^{-44}\,{\rm cm}^2$, and the most conservative limit without making assumptions on the form of the interaction is $\sigma_{\rm DM-p}^{\rm SI}\lesssim 10^{-40}\,{\rm cm}^2$,  obviously much weaker than for concrete scenarios. The dark matter particles from the Virgo Supercluster extend the reach to even larger mass differences, up to $\delta_{\rm DM}\simeq 450$ keV and sets for the isoscalar scenario the limit $\sigma_{\rm DM-p}^{\rm SI}\lesssim 5 \times 10^{-40}\,{\rm cm}^2$ for $\delta_{\rm DM} = 450$ keV; for the interaction mediated by the $Z$-boson the upper limit is $\sigma_{\rm DM-p}^{\rm SI}\lesssim  10^{-41}\,{\rm cm}^2$, while the model independent limit is $\sigma_{\rm DM-p}^{\rm SI}\lesssim 5 \times 10^{-36}\,{\rm cm}^2$.  Similar conclusions apply for the PICO and CRESST experiments, and from the radiopurity measurements on a CaWO$_{4}$ target.

\begin{figure}[t!]
  \centering  
	\includegraphics[width=0.49\textwidth]{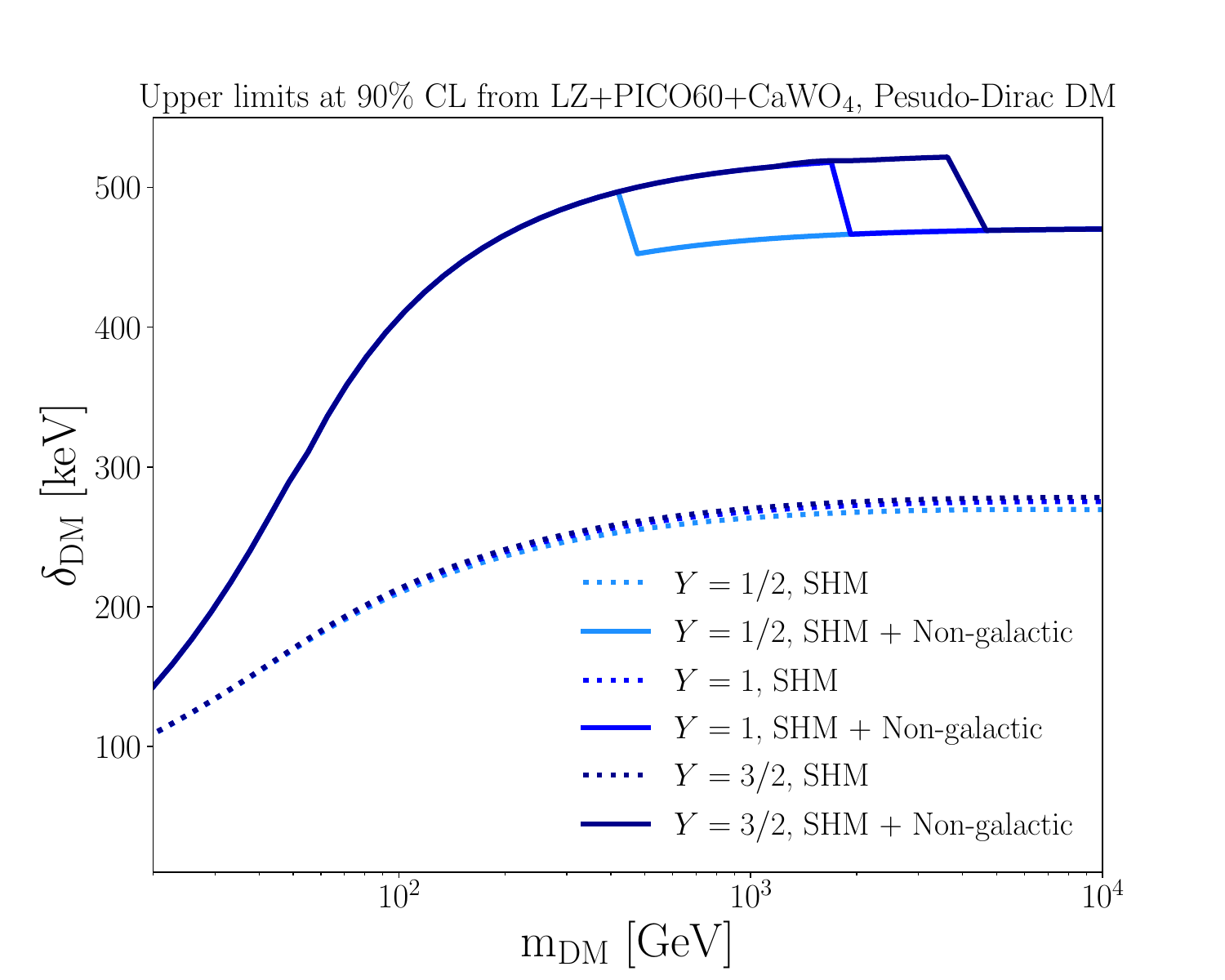}
	\caption{Upper limits on the mass splitting for electroweakly charged (pseudo-)dirac dark matter as a function of the dark matter mass, for different choices of the hypercharge, and including in the flux only the Standard Halo Model component (dotted lines) or also the non-galactic diffuse components (solid lines).}
		\label{fig:delta_vs_mDM}
	\end{figure}

It is interesting to note the complementarity of the different experiments in probing the parameter space of inelastic dark matter scenarios. Both in the scenario of a Majorana dark matter with $f^n=f^p$ and for the scenario with $Z$-boson mediation, LUX-ZEPLIN is the most sensitive probe for small $\delta_{\rm DM}$, whereas the radiopurity measurements on a  CaWO$_{4}$ is the most sensitive probe for large $\delta_{\rm DM}$. PICO-60 is relevant for intermediate values of $\delta_{\rm DM}$, and is in fact the most sensitive current probe of some well motivated dark matter scenarios, as suggested by the gray lines in the Figure, which correspond to the expected cross-section for different scenarios of electroweakly interacting fermionic dark matter.
The complementarity of experiments in probing these scenarios is investigated in Figure \ref{fig:delta_vs_mDM}. The dotted lines show the upper limit on the mass splitting as a function of the dark matter mass assuming the Standard Halo Model. Under this common assumption, LUX-ZEPLIN is the most constraining experiment over the whole parameter space considered. However, when including the non-galactic components, different experiments contribute to set the upper limit, as reflected by the breaks in the solid lines in the Figure:  LUX-ZEPLIN remains as the most sensitive experiment for small dark matter masses, while PICO-60 is the best experiment for larger masses. Further, the dark matter mass at which PICO-60 becomes the leading experiment becomes larger and larger as the dark matter hypercharge increases. As seen in the Figure, for this class of scenarios the non-galactic components in the dark matter flux enhance the sensitivity of experiments to the mass splitting by a factor $\sim 2$ for $m_{\rm DM}=100$ GeV - 1 TeV.

\begin{figure}[t!]
  \centering  
	\includegraphics[width=0.49\textwidth]{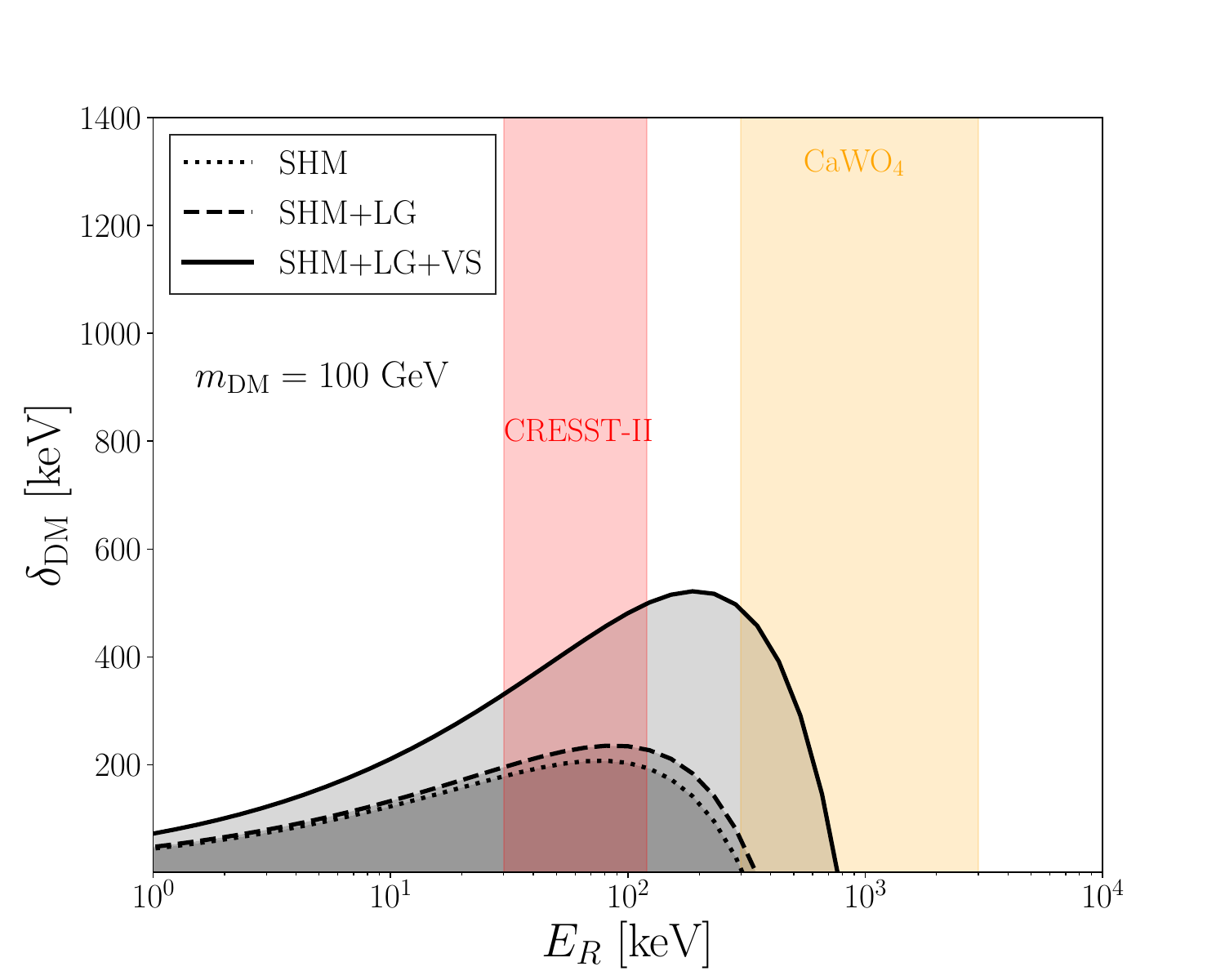}
    \includegraphics[width=0.49\textwidth]{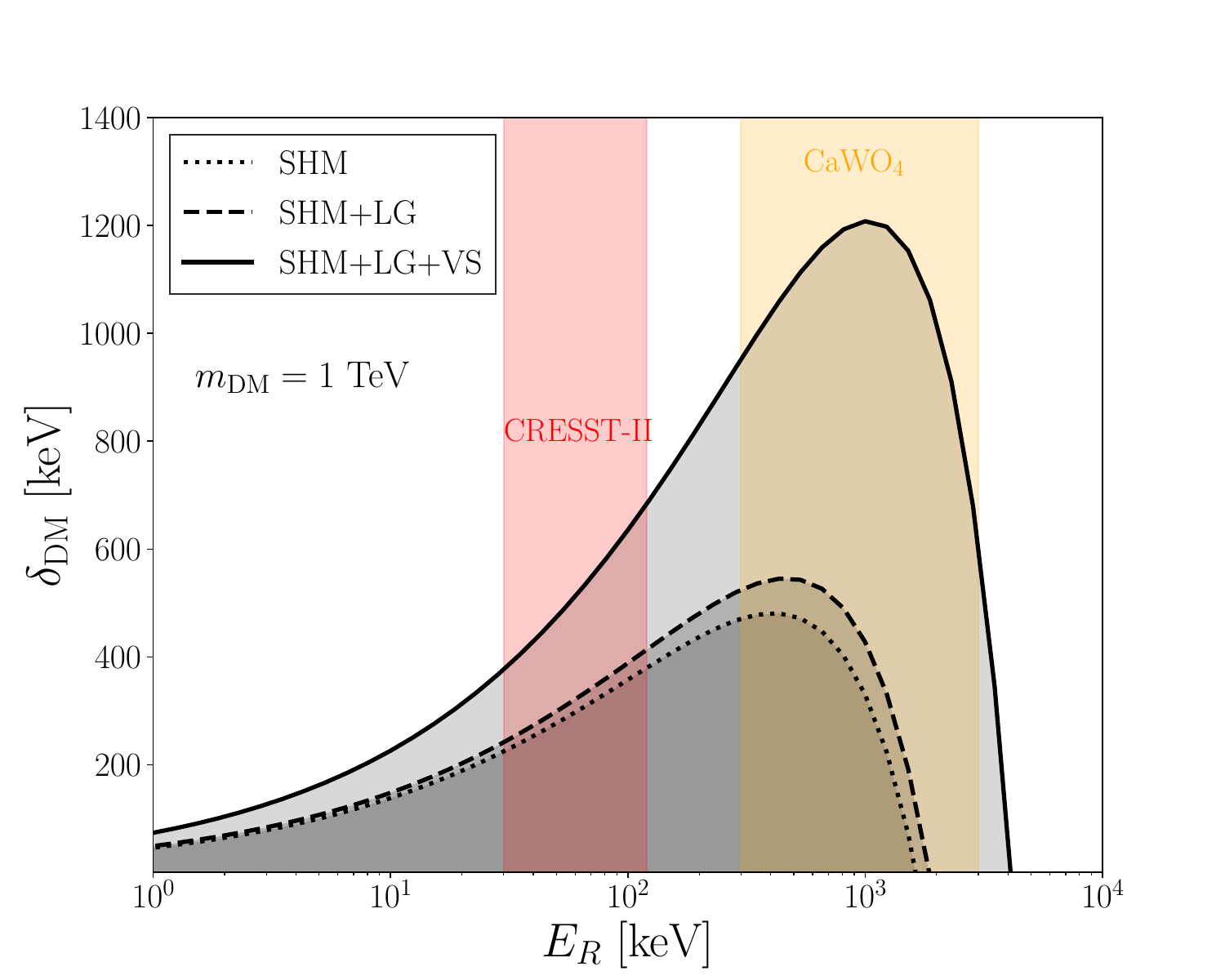} 
	\caption{Values of the mass splitting $\delta_{\rm DM}$  that can produce a recoil energy in a ${}^{184} \rm W$ target for $m_{\rm DM}=100$ GeV (left plot) and $m_{\rm DM}=1$ TeV (right plot) when the maximal velocity of the dark matter particles at Earth is $v_*=764$ km/s (dotted lines), $v_*=820$ km/s (dashed lines) and $v_*=1220$ km/s (solid lines), corresponding respectively to dark matter bound to the Milky Way (described by the Standard Halo Model), bound to the Local Group and bound to the Virgo Supercluster. For comparison, we also show the range of recoil energies that can be detected by the CRESST-II experiment (red band) and by the CaWO$_{4}$ radiopurity measurement (yellow band).}
		\label{fig:delta_vs_ER}
	\end{figure}

It is noteworthy the pivotal role of the radiopurity measurements on a CaWO$_{4}$ target to probe large mass splittings in inelastic dark matter scenarios. This can be understood from the expression for the minimum DM velocity required to induced a recoil with energy $E_R$, Eq.~(\ref{eq:vmin}). Let us consider a velocity distribution where the maximum speed is $v_*$. Then, for an experiment capable of detecting a recoil of a nucleus $A_i$ with energy $E_R$, the maximum mass splitting that can be probed is:
\begin{align}
    \delta_{\rm DM}\leq \sqrt{2 E_R m_{A_i}} v_* -\frac{E_R m_{A_i}}{\mu_{A_i}}\leq \frac{1}{2}\mu_{A_i}v_*^2 \, ,
\end{align}
where the absolute maximum is reached when  $E_R=\mu_{A_i}^2 v_*^2/(2 m_{A_i})$. This is shown in Figure~\ref{fig:delta_vs_ER} for a ${}^{184} \rm W$ target, and for $v_*=764$ km/s, $v_*=820$ km/s, $v_*=1220$ km/s (solid lines), corresponding respectively to the maximal velocity at the Earth of dark matter particles  bound to the Milky Way (described by the Standard Halo Model),  from the Local Group envelope and from the Virgo Supercluster. The plot also shows the range of recoil energies that can be detected by the CRESST-II experiment and by the radiopurity measurements in CaWO$_4$ crystals. As seen in the plot, while CRESST-II can only probe up to $\delta_{\rm DM}\sim 700$ keV, the radiopurity measurements allow to probe up to $\delta_{\rm DM}\sim 1200$ keV, when including the flux component from the dark matter bound to the Virgo Supercluster (however with a lower sensitivity due to the smaller exposure). From this plot it follows that the CRESST experiment would have an enhanced sensitivity to inelastic dark matter scenarios if the window of recoil energies used in the analysis were extended to larger values. Let us note that for low dark matter masses, extending the search window of a given experiment to higher recoil energies would not always help in probing larger values of the mass splitting. This is illustrated in the Figure for $m_{\rm DM}=100$ GeV, from where it is apparent that in order to increase the reach in mass splittings it is necessary to extend the search of the radiopurity CaWO$_4$ measurement to lower recoil energies.

\begin{figure}[t!]
	\centering
		\includegraphics[width=0.45\textwidth]{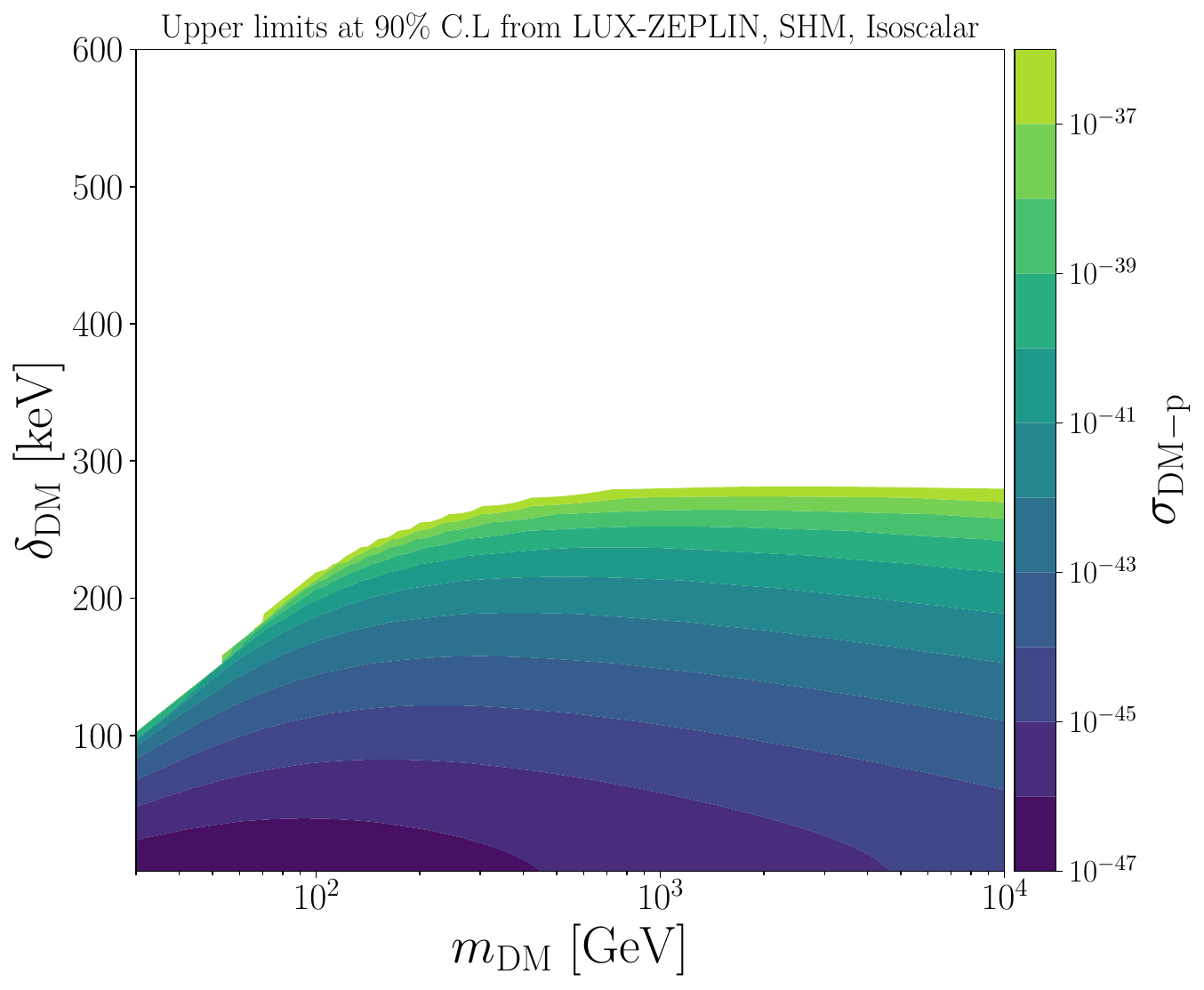}
		\includegraphics[width=0.45\textwidth]{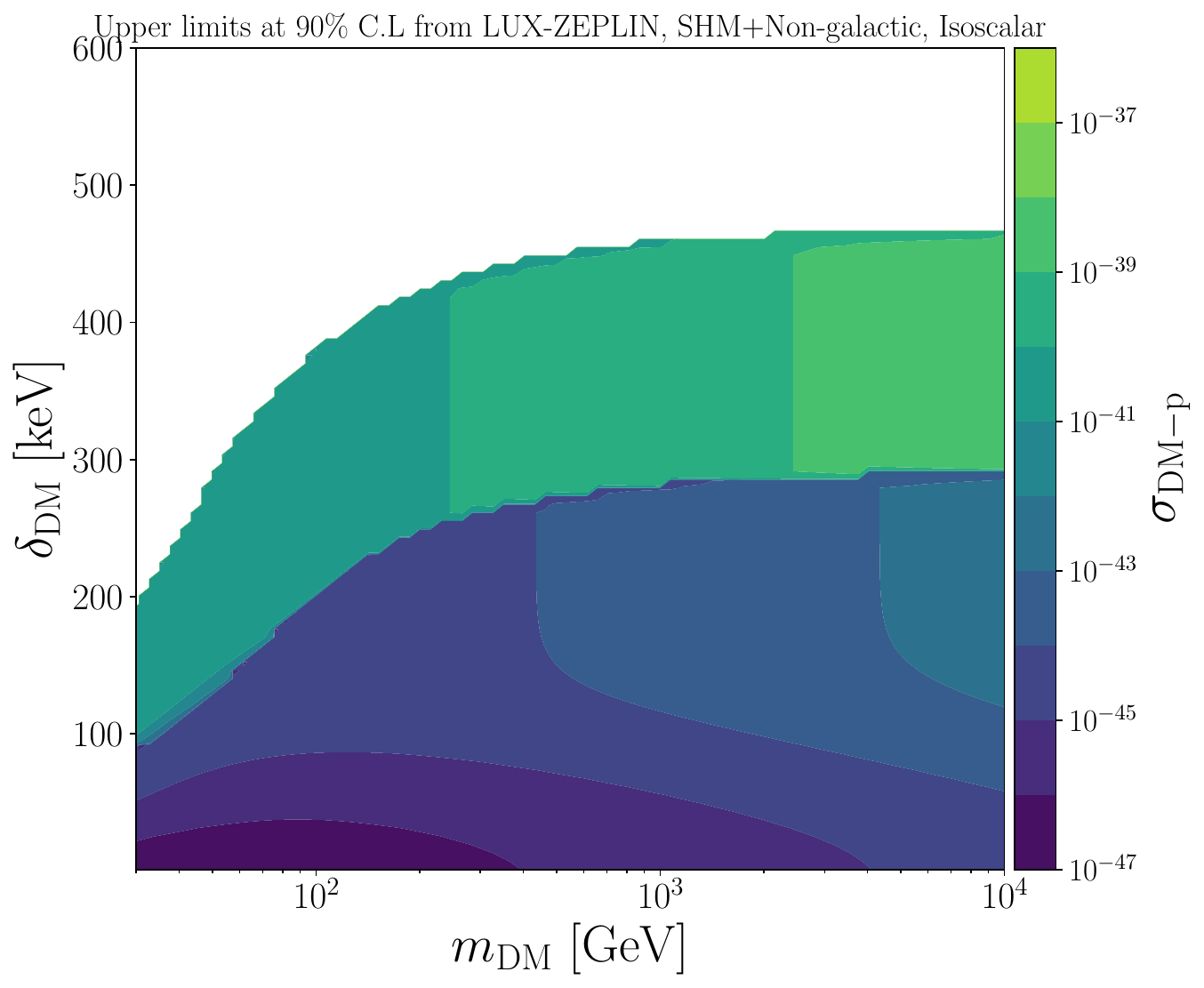} \\
		\includegraphics[width=0.45\textwidth]{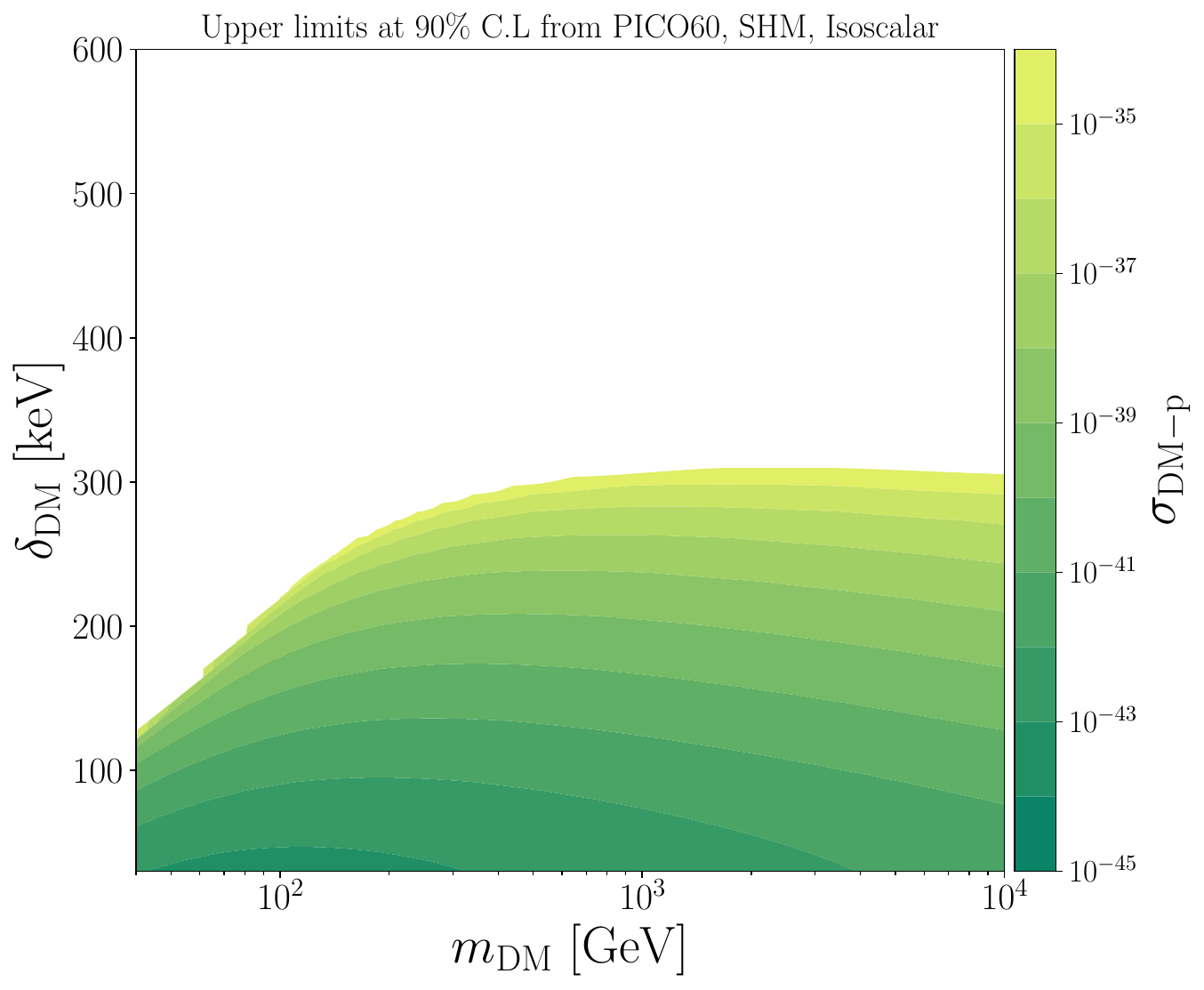}
		\includegraphics[width=0.45\textwidth]{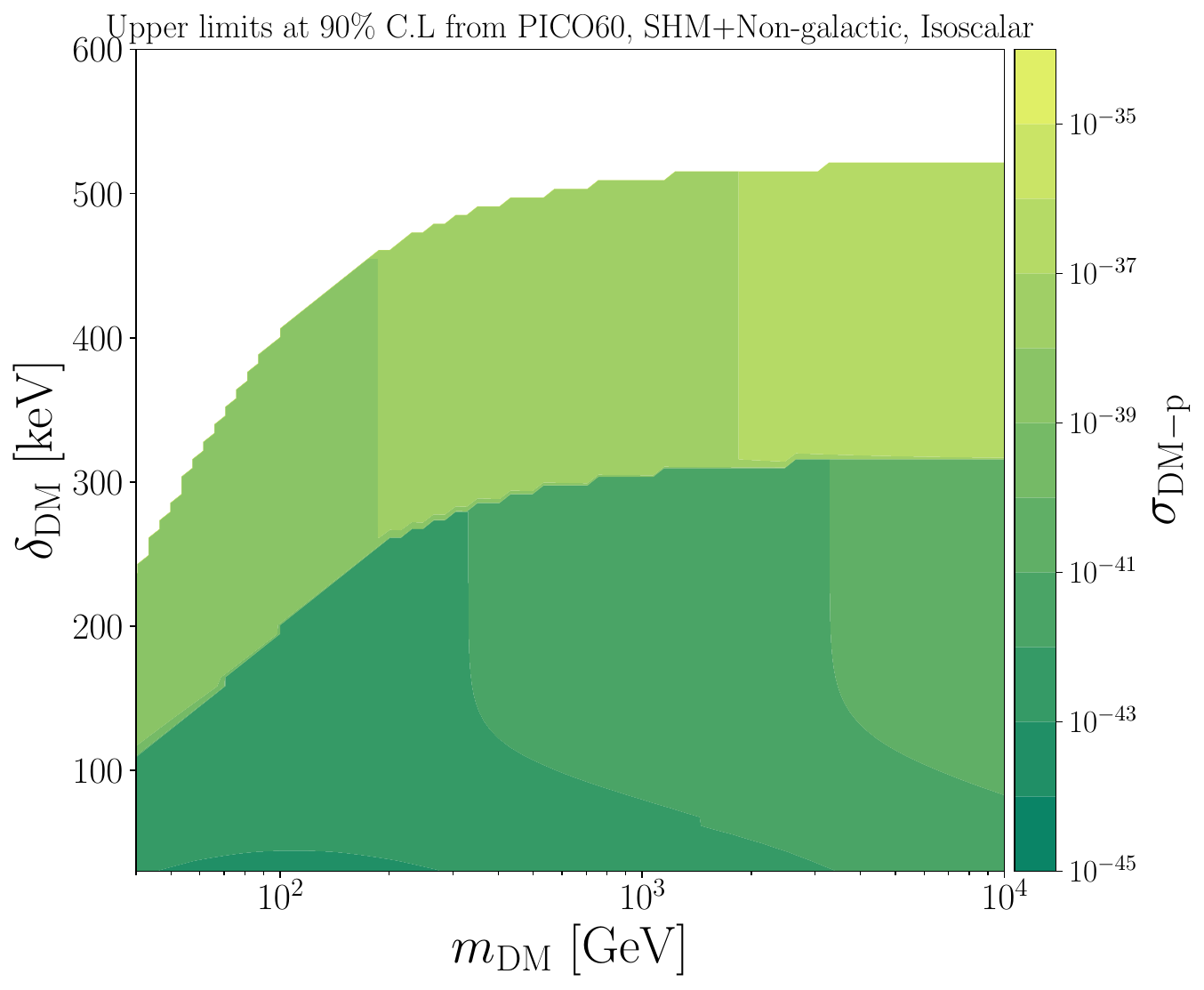} \\
		\includegraphics[width=0.45\textwidth]{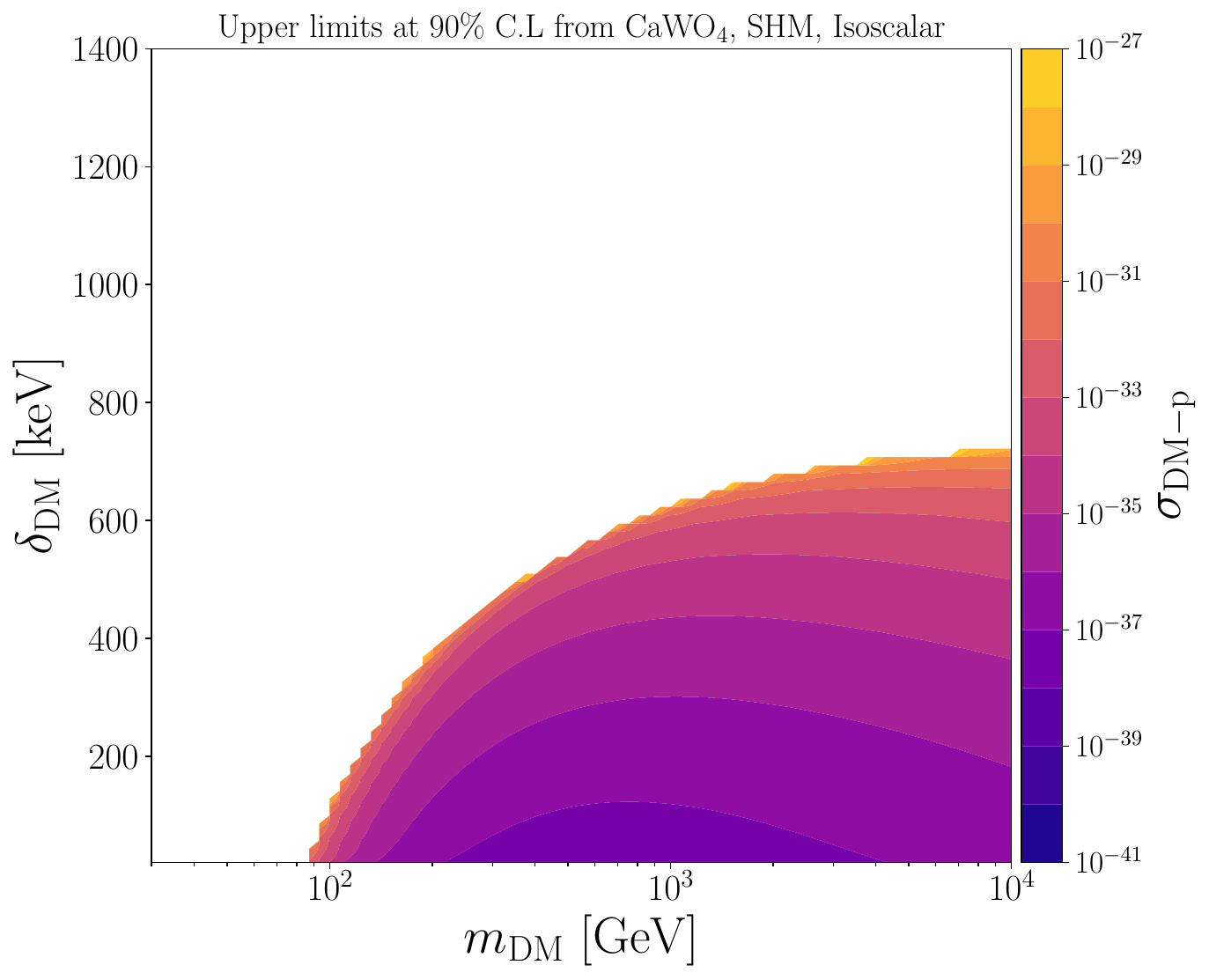}
		\includegraphics[width=0.45\textwidth]{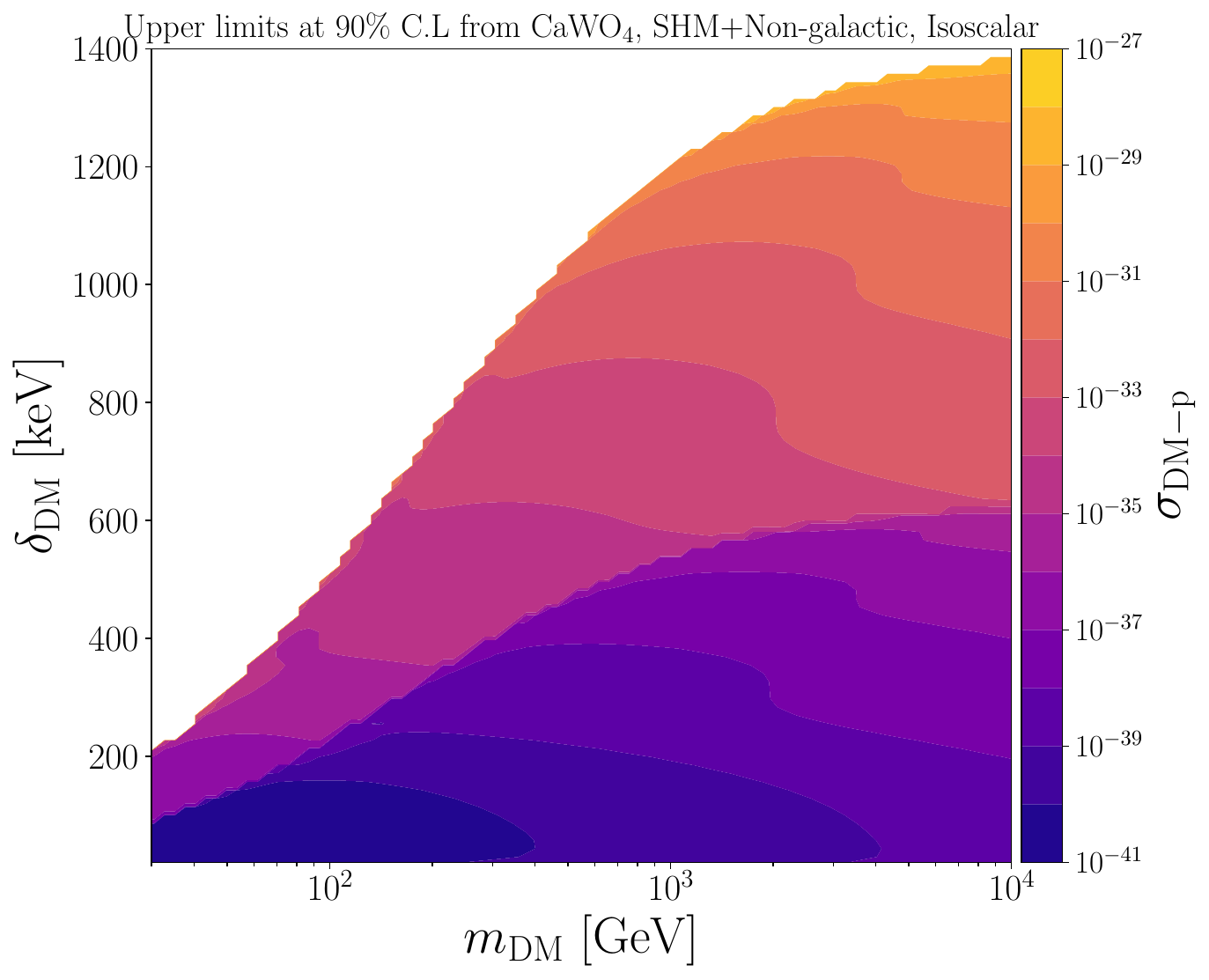}
		\caption{Isocontours of the 90$\%$ C.L. upper limits on the spin-independent dark matter-proton inelastic cross-section for the isoscalar scenario ($f^p=f^n$) in the parameter space spanned by the dark matter mass and mass splitting, from LUX-ZEPLIN (top panels), PICO60 (middle panels) and radiopurity measurements in a CaWO$_{4}$ target (lower panels), assuming that all dark matter in the Solar System is bound to the Milky Way (left panels) or including the non-galactic diffuse component (right panels).}
		\label{fig:2dplots_nuclear}
	\end{figure}

Finally, we show in Figure \ref{fig:2dplots_nuclear} the isocontours with the 90\% C.L. upper limits on the cross-section for different dark matter masses and mass splittings, from LUX-ZEPLIN (top panels), PICO60 (middle panels) and from radiopurity measurements on a CaWO$_{4}$ target (bottom panels), considering that all dark matter in the Solar System is bound to the Milky Way, as commonly assumed (left panels), and including the non-galactic components (right panels). The enhancement in sensitivity is clear from the plots. Further, one can appreciate in the figures a series of ``breaks'', that correspond to those regions in parameter space where the contribution  to the scattering from the Local Group component starts to dominate over the SHM contribution, and to the regions where the contribution from the Virgo Supercluster component starts to dominate over the Local Group contribution. More concretely, if the mass difference is small, the SHM component generates the largest component to the signal rate. However, as the mass difference increases, dark matter particles bound to the Milky Way cannot induce a visible scattering, whereas dark matter particles bound to the Local Group can, thus allowing to probe larger cross-sections (thus resulting in the ``breaks'' in the isocontours in the Figure for certain values of the dark matter mass). The same behaviour occurs for larger mass splittings, when dark matter particles from the Local Group cannot induce detectable recoils, while dark matter particles from the Virgo Supercluster can. Since the fraction of dark matter particles bound to the Virgo Supercluster is rather small,  only $\sim 0.003\%$, the impact of this component is modest, except around the threshold.

	\section{Impact on electron recoils}
	\label{sec:4}
	
	The differential ionization rate induced by dark matter-electron inelastic scattering in liquid xenon, with mass splitting between the two dark matter states given by $\delta_{\rm DM}$, reads:
	\begin{align}
		\frac{dR_{\rm ion}}{d\text{ln}E_{er}}= N_{T}\sum_{n,l}  \int_{v \geq v_{\rm min}^{nl}(E_{er})} \text{d}^{3} v {\mathscr F}(\vec{v}+\vec v_\odot) \, \frac{\text{d} \sigma_{\rm ion}^{nl}}{d\text{ln}E_{er}}(v, E_{er}) \,,
		\label{eq:diff_ionization_rate}
	\end{align}
	where $N_{T}$ is the number of target nuclei and
	\begin{align}
		v_{\rm min}^{nl}(E_{er}) =\sqrt{\frac{2}{m_{\rm DM}}(E_{er}+|E^{nl}|+\delta_{\rm DM})}
		\label{eq:vmin_e}
	\end{align}
	is the minimum dark matter velocity necessary to ionize a bound electron in the $(n,l)$ shell of a xenon atom (with energy  $E^{nl}$), giving a free electron with energy $E_{er}$. Further, $d\sigma_{\rm ion}^{nl} /d\text{ln}E_{er}$ is the differential ionization cross section, given by: 
	\begin{align}
		\frac{d\sigma_{\rm ion}^{nl }}{d\text{ln}  E_{er}}(v,E_{er})=\frac{\bar{\sigma}_{\rm DM-e}}{8 \mu_{\rm DM, e}^{2}v^{2}}\int_{q^{nl}_{\rm min}}^{q^{nl}_{\rm max}} dq q \left |f_{\rm ion}^{nl}(k',q)  \right |^{2} \left |F_{\mathrm{DM}}(q)  \right |^{2}.
	\end{align}
	Here, $\mu_{\rm DM,e}$ is the reduced mass of the dark matter-electron system, 
	$\bar{\sigma}_{\rm DM-e}$ is the dark matter-free electron scattering cross section at fixed momentum transfer $q=\alpha m_{e}$, $\left|f_{ion}^{nl}(k',q)  \right |^{2}$ is the ionization form factor of an electron in the $(n,l)$ shell with final momentum $k'=\sqrt{2m_{e}E_{er}}$ and momentum transfer $q$, and $F_{\mathrm{DM}}(q)$ is a form factor that encodes the $q$-dependence of the squared matrix element for dark matter-electron scattering and depends on the mediator under consideration. The maximum and minimum values of the momentum transfer needed to ionize a bound electron in the $(n,l)$ shell recoil with energy $E_{er}$ from the interaction of a dark matter particle with speed $v$ are:
	\begin{align}
		q^{nl}_{\substack{{\rm max}\\{\rm min}}}(E_{er})= m_{\rm DM} v \left[1\pm\sqrt{1-\left(\frac{v_{\rm min}^{nl}(E_{er})}{v}\right)^2}\right],
		\label{eq:q_max_min}
	\end{align}
	with $v_{\rm min}^{nl}(E_{er})$ defined in Eq.~(\ref{eq:vmin_e}). 
	Finally, the total number of expected ionization events reads ${\cal N}=R_{ion}\cdot \mathcal{E}$,  with 
	$R_{ion}$ the total ionization rate, calculated from integrating Eq.(\ref{eq:diff_ionization_rate}) over the experimentally measured recoil energies, and $\mathcal{E}$ the exposure ({\it i.e.} mass multiplied by live-time) of the experiment.
	
	In semiconductor detectors, the electron excitation rate induced by dark matter-electron inelastic scatterings, with a mass splitting $\delta_{\rm DM}$, reads \cite{Knapen_2017, Knapen_2022}
\begin{align}
				R=&\frac{1}{\rho_{T}} \frac{\bar{\sigma}_{\rm DM-e}}{\mu_{\rm DM,e}^{2}} \frac{\pi}{\alpha} \int d^{3} v \frac{\mathscr F(\vec{v}+\vec v_\odot)}{v} \int \frac{d^{3} q}{(2 \pi)^{3}} q^{2}\left|F_{\rm DM}(q)\right|^{2} \nonumber 
				\\&\int \frac{d \omega}{2 \pi} \frac{1}{1-e^{-\beta \omega}} \operatorname{Im}\left[\frac{-1}{\epsilon(\omega, \vec q)}\right] \delta\left(\omega+\delta_{\rm DM}+\frac{q^{2}}{2 m_{\chi}}-\vec{q} \cdot \vec{v}\right) \, ,
			\end{align}
where $w$ is the energy deposited in the material, $\vec q$ is the momentum transfer of the process, and $\rho_{T}$ is the target density. The rate involves an integration of the Electronic Loss Function (ELF) of the target material, which we calculate with \texttt{DarkELF} \cite{Knapen_2022}. For the dielectric function $\epsilon(\omega, \mathbf{q})$, we use the Lindhard method, which treats the target as a non-interacting Fermi liquid. Finally, the total number of events reads ${\cal N}=R \cdot \mathcal{E}$, with $\mathcal{E}$ the exposure ({\it i.e.} mass multiplied by live-time) of the experiment.
	
\begin{figure}[t!]
 		\includegraphics[width=0.5\textwidth]{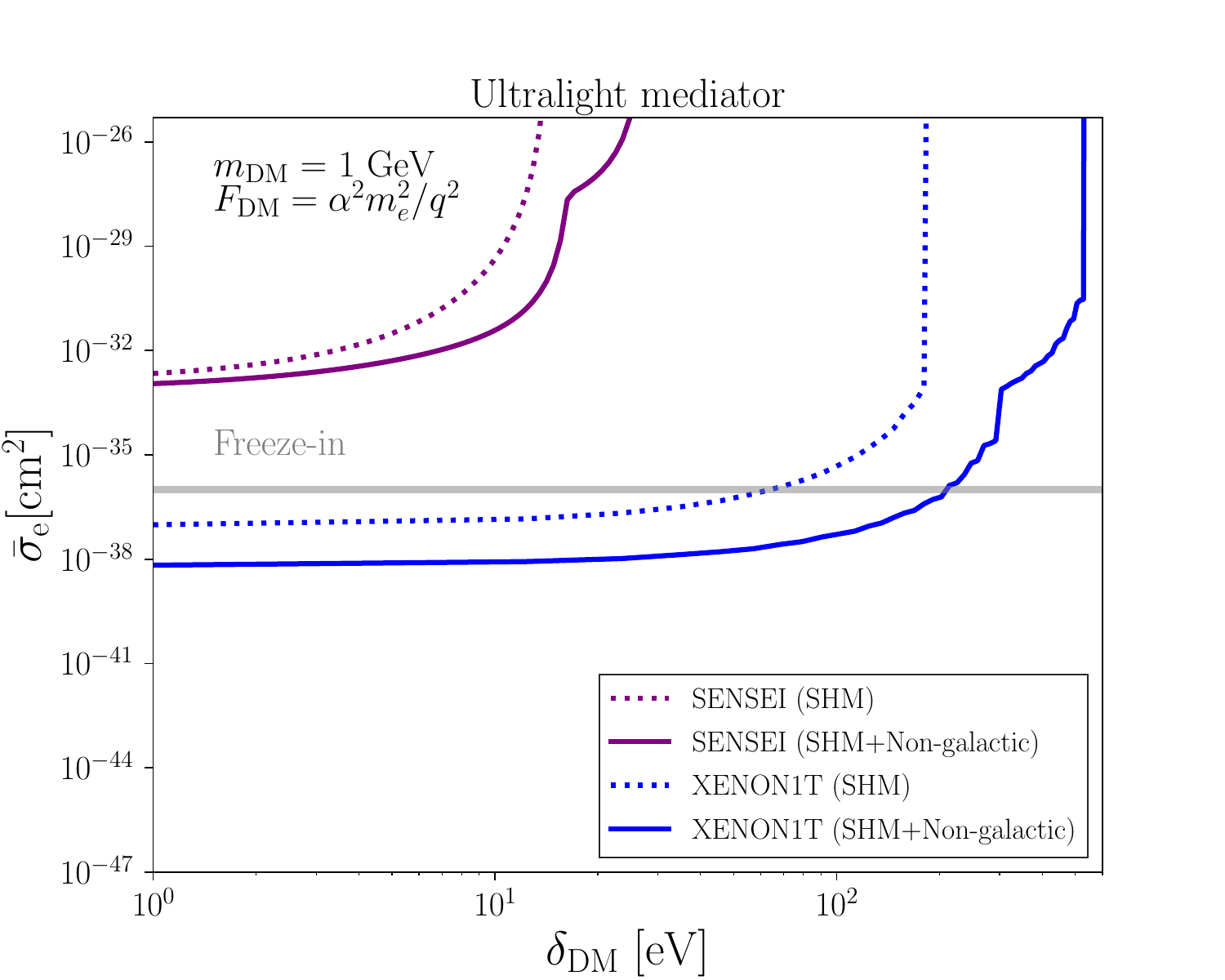}
        \includegraphics[width=0.5\textwidth]{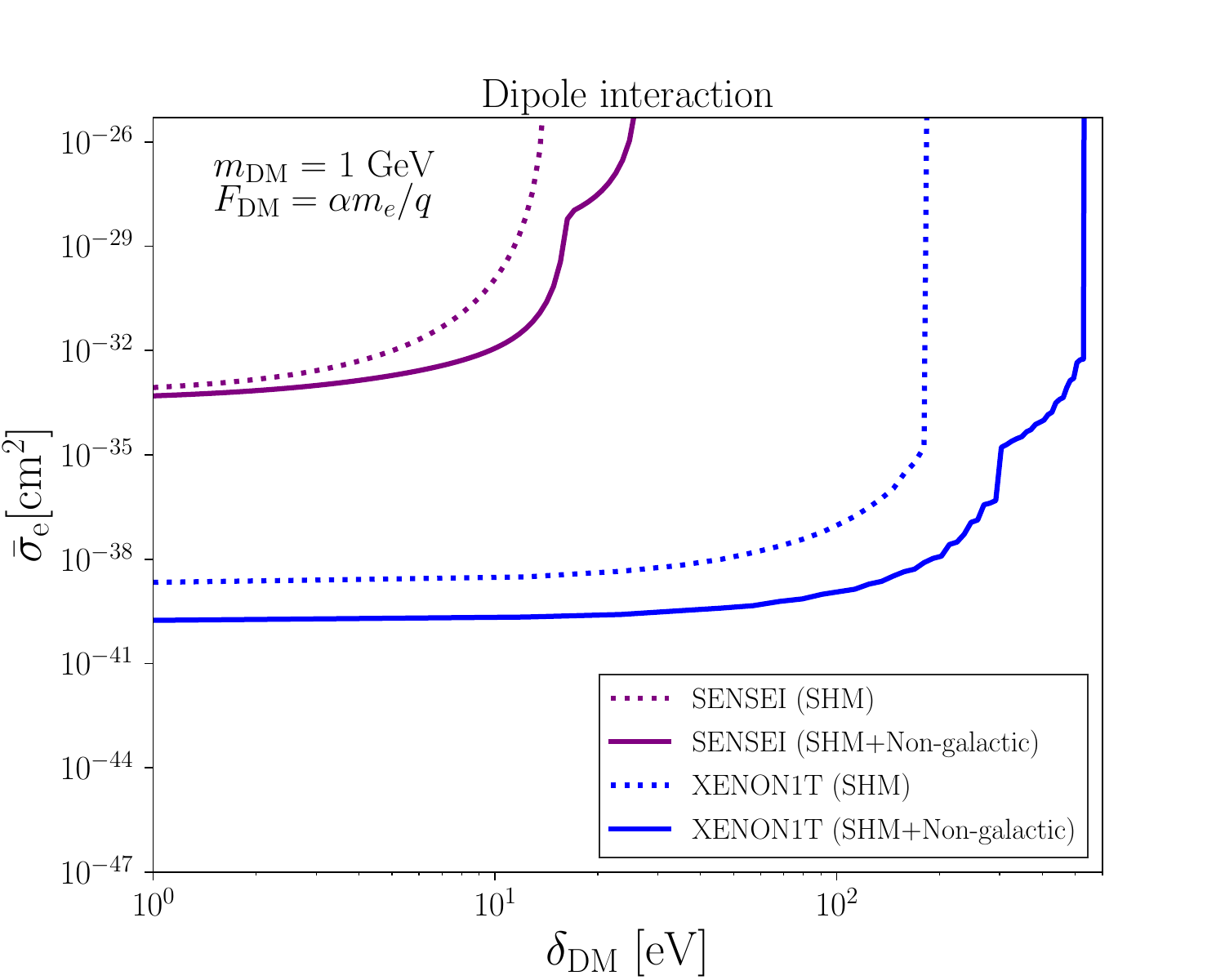}
        \begin{center}
        \includegraphics[width=0.5\textwidth]{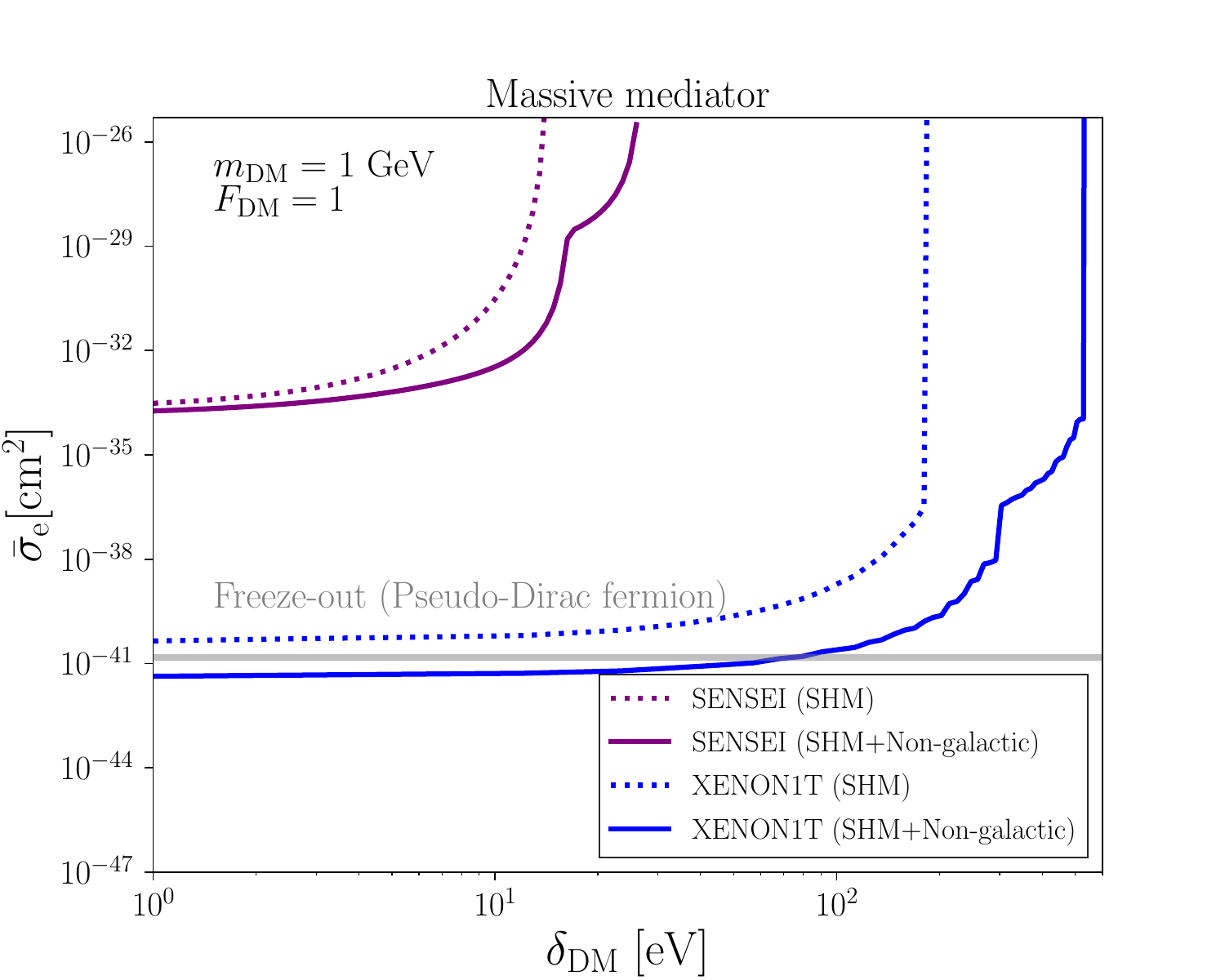}
        \end{center}
		\caption{90$\%$ C.L upper limits on the spin-independent dark matter-electron inelastic cross section for a dark matter mass of 1 GeV, as a function of the mass splitting, from XENON1T (blue) and SENSEI (purple), when the dark matter-electron interaction is mediated by an ultralight dark photon (upper left plot), by a dipole operator (upper right plot), or by a heavy mediator (lower plot).}
		\label{fig:upper_limits_electron_recoils}
	\end{figure}

The non-observation of a significant excess of electron recoils in a given experiment allows to set upper limits on the dark matter-electron scattering cross section, for a given dark matter mass and a given mass splitting between the dark matter particle and the heavier neutral state. We show in Figure \ref{fig:upper_limits_electron_recoils},  upper limits on the inelastic dark matter-electron cross section versus mass splitting for a fixed dark matter mass of $m_{\rm DM}=1$ GeV from XENON1T~\cite{Aprile_2018}(blue lines), and from the semiconductor experiment SENSEI~\cite{Barak_2020}(purple lines), both when considering the SHM flux only (solid lines), and when including the non-galactic components to the dark matter flux (dotted lines). In the upper plots, we take the form factor $F_{\rm DM}=\alpha^2m_e^2/q^2$, corresponding to an ultralight or massless mediator. In the middle plots, we take the form factor $F_{\rm DM}=\alpha m_e/q$, corresponding to an electric dipole interaction, and in the lower plots we take the form factor $F_{\rm DM}=1$, corresponding to a heavy mediator \cite{Essig:2022dfa, Catena_2020}. 
	
	As can be seen in the Figure, the non-galactic components enhance the sensitivity  to the mass splitting of both XENON1T and SENSEI by a factor of $\sim$ 2, compared to the sensitivity estimated from considering just the galactic component. This conclusion holds independently of the choice of the dark matter form factor. 
	Further, the reach in cross-section is enhanced due to the non-galactic components, especially at low mass splittings, being the effect stronger for XENON1T than for SENSEI. For comparison, we also show as a grey band the cross section for which the observed dark matter abundance is reproduced via freeze-in in the case of an ultralight mediator~\cite{Essig:2017kqs}, or via freeze-out in the case of a heavy mediator~\cite{CarrilloGonzalez:2021lxm}. Clearly, the non-galactic dark matter components allow to probe larger values of the mass splitting. Finally, we also show in Figure 7 the isocontours with the 90\% C.L. upper limits on the dark matter-electron scattering cross-section for different dark matter masses and mass splittings, from SENSEI (upper panels) and XENON1T (lower panels), considering that all dark matter in the Solar System is bound to the Milky Way (left panels), and including the non-galactic components (right panels). The non-galactic components enhances the reach in mass splittings by a factor or $\sim 1.5$ for SENSEI and $\sim 2.5$ for XENON1T, allowing to probe lower dark matter masses and cross sections in both cases.

 	\begin{figure}[t!]
	\centering
		\includegraphics[width=0.45\textwidth]{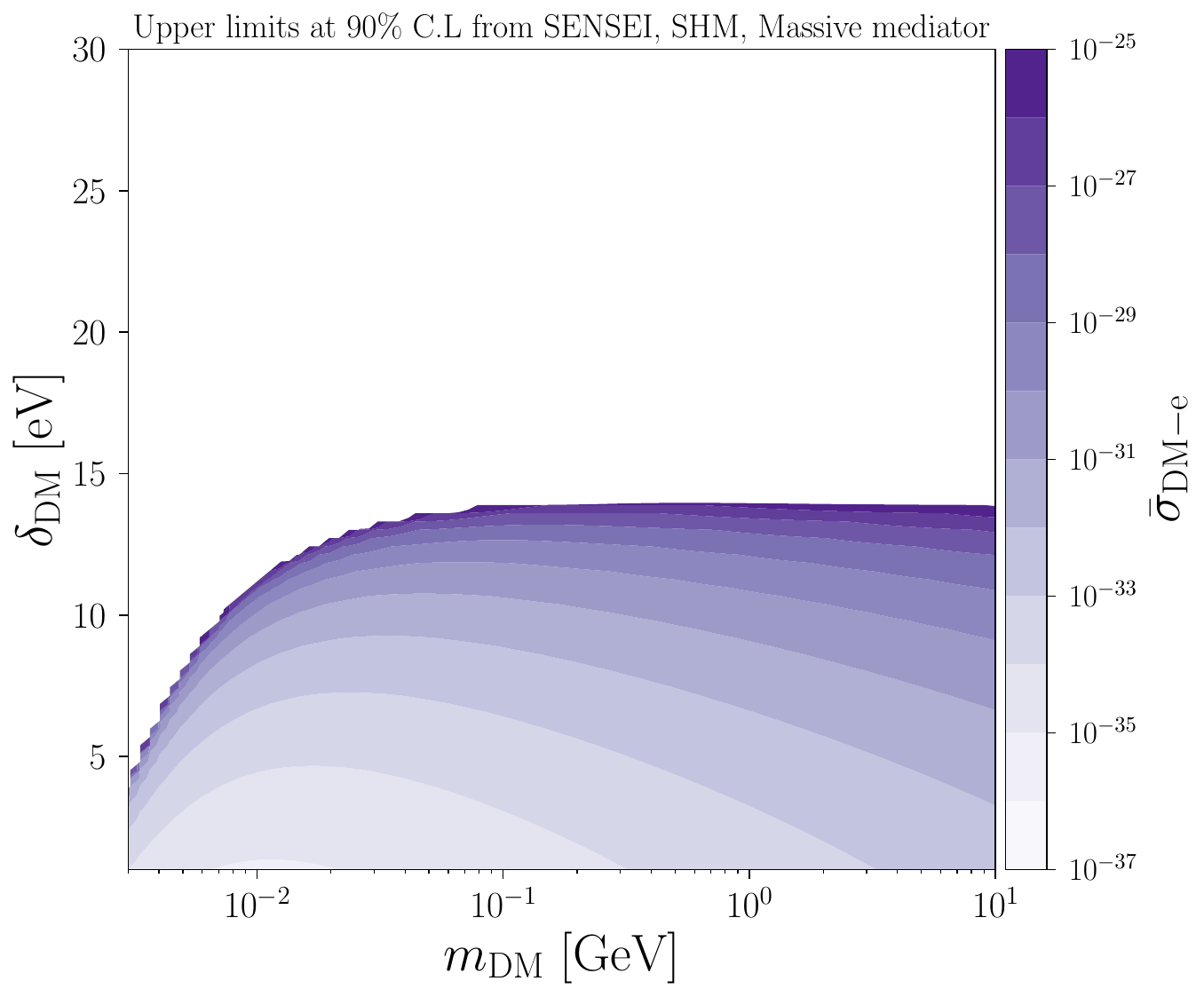}
		\includegraphics[width=0.45\textwidth]{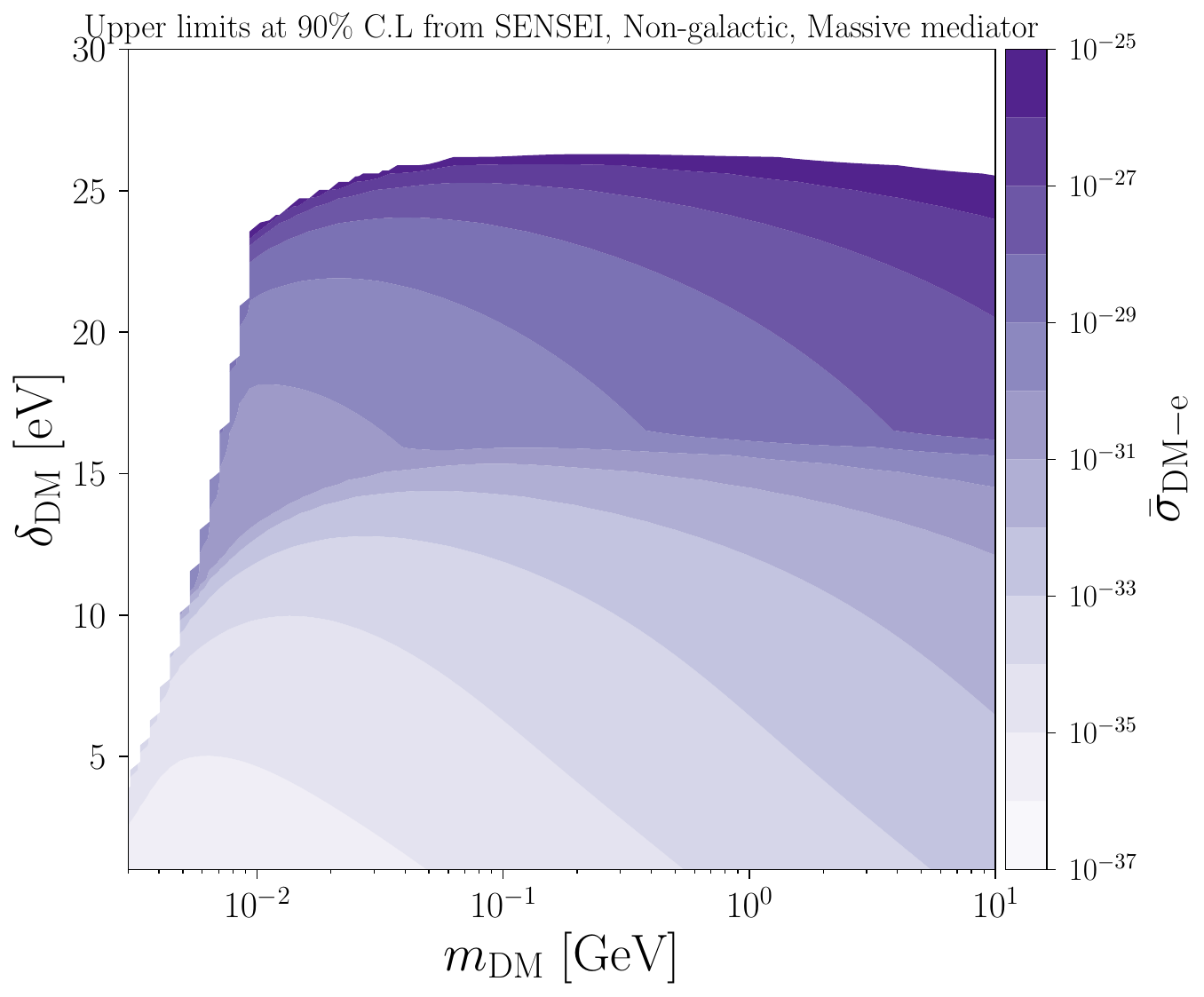}
		\includegraphics[width=0.45\textwidth]{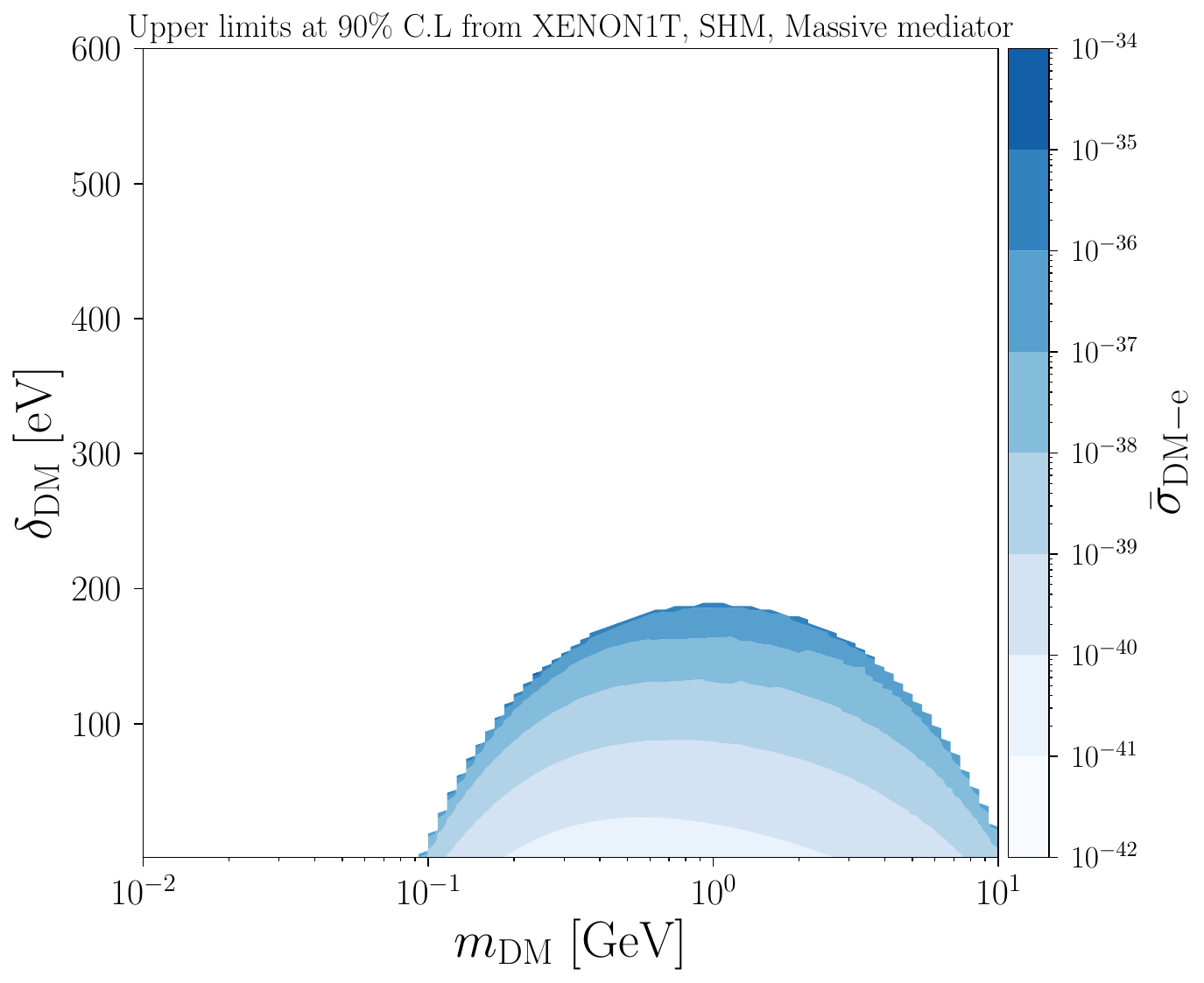}
		\includegraphics[width=0.45\textwidth]{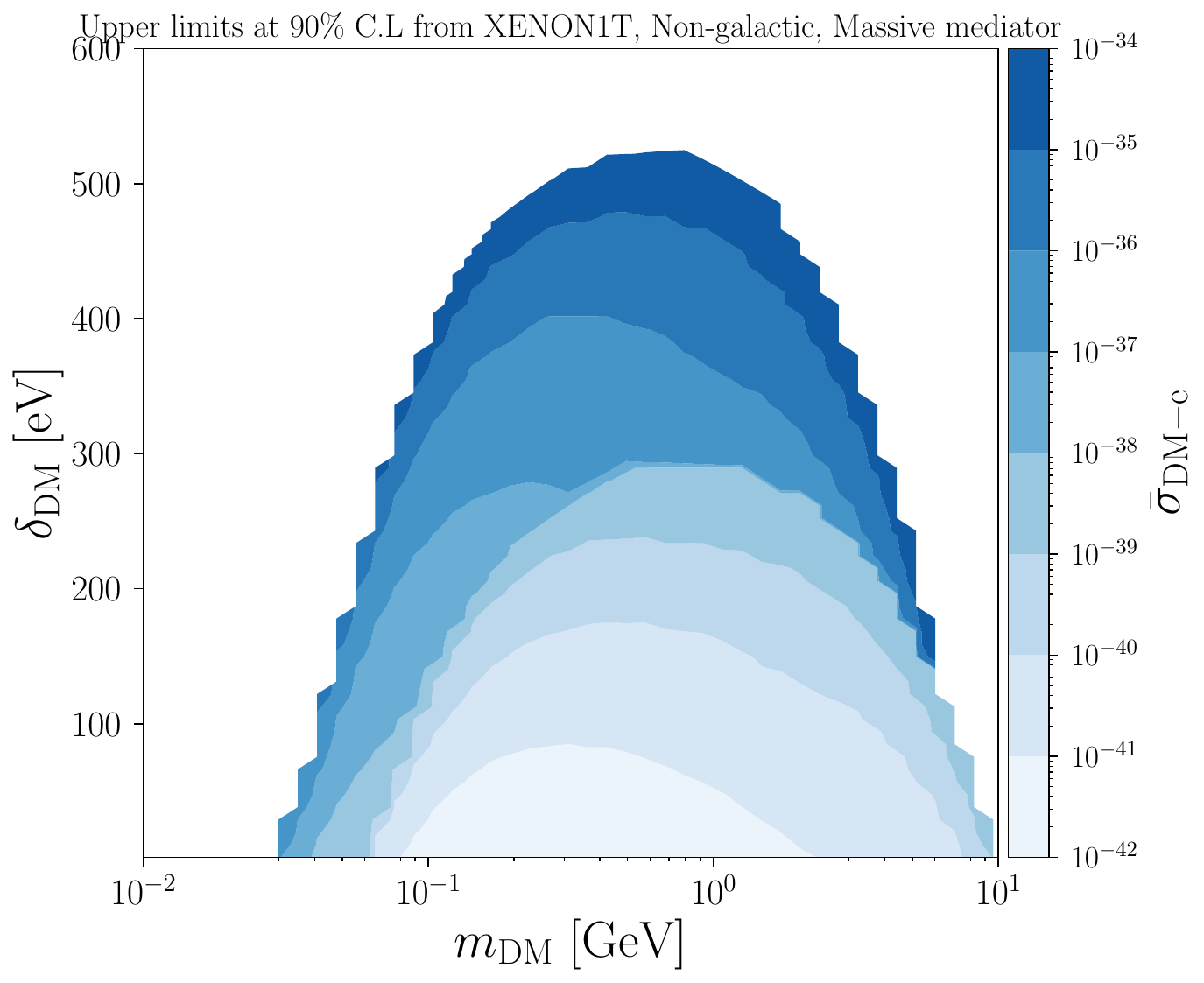}
		\caption{Isocontours of the 90$\%$ C.L. upper limits on the dark matter-electron inelastic scattering cross-section for the heavy mediator scenario ($F_{\rm DM}=1$) in the parameter space spanned by the dark matter mass and mass splitting, from SENSEI (top panels), and XENON1T (lower panels), assuming that all dark matter in the Solar System is bound to the Milky Way (left panels) or including the non-galactic component diffuse (right panels).}
		\label{fig:2dplots_electron}
	\end{figure}

	\section{Conclusions}\label{sec:conclusions}\label{sec:5}
	
	We have investigated the impact of a non-galactic diffuse dark matter component inside the Solar System for the detection of the inelastic scattering of a dark matter particle in direct search experiments. Concretely, we have considered the contribution to the dark matter flux from dark matter particles in the envelope of the Local Group and from the Virgo Supercluster. Their speeds in the galactic frame are $\sim 600$ km/s and $\sim 1000$ km/s, respectively, which are larger than the maximal speed of dark matter particles bound to the Milky Way, $\sim 540$ km/s. As a result, the region of parameter space that can be probed with current experiments is larger than reported in previous works, that implicitly assumed that the Milky Way is an isolated galaxy in the Universe. 
	
	For nuclear recoils, the non-galactic component expands the reach in mass splitting at the LUX-ZEPLIN, PICO60, and CRESST-II experiments by a factor $\sim 2$ in the mass range $m_{\rm DM}=10$ GeV- 10 TeV, and enhances significantly the reach in cross-section, especially close to the kinematic threshold for the galactic dark matter. For instance, for $m_{\rm DM}=1$ TeV and $\delta_{\rm DM}=250$ keV, the sensitivity to the cross-section improves by about three orders of magnitude. We have also stressed the relevance of experiments capable of detecting high recoil energies for probing the parameter space of inelastic dark matter scenarios. We have illustrated this capability with the radiopurity measurements in CaWO$_4$ crystals performed by the CRESST collaboration, and which allows to probe up to $\delta_{\rm DM}\sim 1.2$ MeV (1.4 MeV) for $m_{\rm DM}=1$ TeV (10 TeV).   For electron recoils, the conclusions are analogous, allowing to increase reach in mass splitting of the XENON1T and SENSEI experiments also by a factor $\sim 2$ for dark matter masses in the range  $m_{\rm DM}=0.01$ GeV-10 GeV,

\subsection*{Acknowledgments}

The work of GH and AI was supported by the Collaborative Research Center SFB1258 and by the Deutsche Forschungsgemeinschaft (DFG, German Research Foundation) under Germany's Excellence
Strategy - EXC-2094 - 390783311. 
The work of SS is supported by Grant-in-Aid for Scientific Research from the Ministry of Education, Culture, Sports, Science, and Technology (MEXT), Japan, 18K13535, 20H01895, 20H05860 and 21H00067, and by World Premier International Research Center Initiative (WPI), MEXT, Japan. 
	
	\appendix
	\section{Derivation of upper limits from direct detection experiments}
	\label{sec:details}
	To derive upper limits on the inelastic dark matter-nucleon scattering cross section, as a function of the dark matter mass and/or the dark matter mass splitting, we follow a poissonian-likelihood approach, and we calculate the rates for the different experiments/detectors independently. For the LUX-ZEPLIN experiment, we use the data from \cite{LZ:2022ufs}, with an exposure of 0.904 tonne$\times$year, a region of interest extending from 2 keV to 70 keV, and the efficiency function reported by the collaboration. Given the agreement of the number of signal events with the background prediction reported by the collaboration, we take a $90\%$ C.L. upper limit on the number of signal events of 2.71. For the PICO-60 experiment, we use the results from \cite{Amole_2016}, corresponding to an exposure of 9.356 kg$\times$year, a region of interest extending from 13.5 keV to 100 keV, and the efficiency function reported by the collaboration. Since PICO-60 observed no signal events, we take a $90\%$ C.L. upper limit on the number of signal events of 2.71. For CRESST-II, we use the published data \cite{Angloher_2016}, corresponding to an exposure of 52 kg$\times$days. We do not consider as signal events those belonging to the acceptance region of the experiment at low recoil energies, but instead, we consider the recoil energy region extending from 30 keV to 120 keV, which gives an upper limit of 4 signal events. Finally, for the CaWO$_{4}$ radiopurity measurement from \cite{M_nster_2014}, we take an exposure of 90.10 kg$\times$days, with a recoil energy region extending from 300 keV to 2000 keV, and a number of 3 signal events.
	
	For the inelastic dark matter-electron scattering cross-section, we derive upper limits at 90$\%$ C.L  at fixed momentum transfer $q=\alpha m_{e}$ using data from XENON1T~\cite{Aprile_2018} and SENSEI~\cite{Barak_2020}. We consider the observed event rate XENON1T between 150-3000 photoelectrons (PE), which corresponds to the range 0.18 keV$_{ee}$ to 3.5 keV$_{ee}$ (kiloelectronvolt electron equivalent). We take the efficiency function from \cite{Aprile_2018}, an exposure of 22 $\pm$ 3 tonne-days and an upper limit on the number of events of 39.2. For SENSEI, we sum-up the observed events in the energy bins ranging from 4.91 eV to 16.31 eV, resulting in an upper limit of 4.957 events per gram day of exposure. Further, we use the efficiency reported by the collaboration in every energy bin \cite{Barak_2020}.
	
	\printbibliography

@article{Bertone:2004pz,
    author = "Bertone, Gianfranco and Hooper, Dan and Silk, Joseph",
    title = "{Particle dark matter: Evidence, candidates and constraints}",
    eprint = "hep-ph/0404175",
    archivePrefix = "arXiv",
    reportNumber = "FERMILAB-PUB-04-047-A",
    doi = "10.1016/j.physrep.2004.08.031",
    journal = "Phys. Rept.",
    volume = "405",
    pages = "279--390",
    year = "2005"
}

@article{Feng:2010gw,
    author = "Feng, Jonathan L.",
    title = "{Dark Matter Candidates from Particle Physics and Methods of Detection}",
    eprint = "1003.0904",
    archivePrefix = "arXiv",
    primaryClass = "astro-ph.CO",
    reportNumber = "UCI-TR-2009-13",
    doi = "10.1146/annurev-astro-082708-101659",
    journal = "Ann. Rev. Astron. Astrophys.",
    volume = "48",
    pages = "495--545",
    year = "2010"
}

@article{Jungman:1995df,
    author = "Jungman, Gerard and Kamionkowski, Marc and Griest, Kim",
    title = "{Supersymmetric dark matter}",
    eprint = "hep-ph/9506380",
    archivePrefix = "arXiv",
    reportNumber = "SU-4240-605, UCSD-PTH-95-02, IASSNS-HEP-95-14, CU-TP-677",
    doi = "10.1016/0370-1573(95)00058-5",
    journal = "Phys. Rept.",
    volume = "267",
    pages = "195--373",
    year = "1996"
}

@article{Bergstrom:2000pn,
    author = {Bergstr\"om, Lars},
    title = "{Nonbaryonic dark matter: Observational evidence and detection methods}",
    eprint = "hep-ph/0002126",
    archivePrefix = "arXiv",
    doi = "10.1088/0034-4885/63/5/2r3",
    journal = "Rept. Prog. Phys.",
    volume = "63",
    pages = "793",
    year = "2000"
}

@article{Bernabei:2007gr,
    author = "Bernabei, R. and others",
    title = "{Investigating electron interacting dark matter}",
    eprint = "0712.0562",
    archivePrefix = "arXiv",
    primaryClass = "astro-ph",
    doi = "10.1103/PhysRevD.77.023506",
    journal = "Phys. Rev. D",
    volume = "77",
    pages = "023506",
    year = "2008"
}

@article{Lewin:1995rx,
    author = "Lewin, J. D. and Smith, P. F.",
    title = "{Review of mathematics, numerical factors, and corrections for dark matter experiments based on elastic nuclear recoil}",
    reportNumber = "RAL-TR-95-024",
    doi = "10.1016/S0927-6505(96)00047-3",
    journal = "Astropart. Phys.",
    volume = "6",
    pages = "87--112",
    year = "1996"
}

@article{MarrodanUndagoitia:2015veg,
    author = "Marrod\'an Undagoitia, Teresa and Rauch, Ludwig",
    title = "{Dark matter direct-detection experiments}",
    eprint = "1509.08767",
    archivePrefix = "arXiv",
    primaryClass = "physics.ins-det",
    doi = "10.1088/0954-3899/43/1/013001",
    journal = "J. Phys. G",
    volume = "43",
    number = "1",
    pages = "013001",
    year = "2016"
}

@article{Belanger:2008sj,
    author = "Belanger, G. and Boudjema, F. and Pukhov, A. and Semenov, A.",
    title = "{Dark matter direct detection rate in a generic model with micrOMEGAs 2.2}",
    eprint = "0803.2360",
    archivePrefix = "arXiv",
    primaryClass = "hep-ph",
    reportNumber = "LAPTH-1237-08",
    doi = "10.1016/j.cpc.2008.11.019",
    journal = "Comput. Phys. Commun.",
    volume = "180",
    pages = "747--767",
    year = "2009"
}

@article{Cerdeno:2010jj,
    author = "Cerdeno, David G. and Green, Anne M.",
    title = "{Direct detection of WIMPs}",
    eprint = "1002.1912",
    archivePrefix = "arXiv",
    primaryClass = "astro-ph.CO",
    reportNumber = "FTUAM-10-02, IFT-UAM-CSIC-10-07, FTUAM-10-02; IFT-UAM/CSIC-10-07",
    doi = "10.1017/CBO9780511770739.018",
    pages = "347--369",
    month = "2",
    year = "2010"
}

@article{Goodman:1984dc,
    author = "Goodman, Mark W. and Witten, Edward",
    editor = "Srednicki, M. A.",
    title = "{Detectability of Certain Dark Matter Candidates}",
    reportNumber = "Print-85-0030 (PRINCETON)",
    doi = "10.1103/PhysRevD.31.3059",
    journal = "Phys. Rev. D",
    volume = "31",
    pages = "3059",
    year = "1985"
}

@article{Nieves:1981zt,
    author = "Nieves, Jose F.",
    title = "{Electromagnetic Properties of Majorana Neutrinos}",
    reportNumber = "Print-81-0796 (PUERTO RICO)",
    doi = "10.1103/PhysRevD.26.3152",
    journal = "Phys. Rev. D",
    volume = "26",
    pages = "3152",
    year = "1982"
}

@article{Cirelli:2005uq,
    author = "Cirelli, Marco and Fornengo, Nicolao and Strumia, Alessandro",
    title = "{Minimal dark matter}",
    eprint = "hep-ph/0512090",
    archivePrefix = "arXiv",
    reportNumber = "DFTT40-2005, IFUP-TH-2005-34",
    doi = "10.1016/j.nuclphysb.2006.07.012",
    journal = "Nucl. Phys. B",
    volume = "753",
    pages = "178--194",
    year = "2006"
}

@article{Nagata_2015,
	doi = {10.1007/jhep01(2015)029},
  
	url = {https://doi.org/10.1007%2Fjhep01%282015%29029},
  
	year = 2015,
	month = {jan},
  
	publisher = {Springer Science and Business Media {LLC}
},
  
	volume = {2015},
  
	number = {1},
  
	author = {Natsumi Nagata and Satoshi Shirai},
  
	title = {Higgsino dark matter in high-scale supersymmetry},
  
	journal = {Journal of High Energy Physics}
}

@article{Alves_2010,
	doi = {10.1016/j.physletb.2010.08.006},
  
	url = {https://doi.org/10.1016%2Fj.physletb.2010.08.006},
  
	year = 2010,
	month = {sep},
  
	publisher = {Elsevier {BV}
},
  
	volume = {692},
  
	number = {5},
  
	pages = {323--326},
  
	author = {Daniele S.M. Alves and Siavosh R. Behbahani and Philip Schuster and Jay G. Wacker},
  
	title = {Composite inelastic dark matter},
  
	journal = {Physics Letters B}
}

@article{Knapen_2017,
	doi = {10.1103/physrevd.96.115021},
  
	url = {https://doi.org/10.1103%2Fphysrevd.96.115021},
  
	year = 2017,
	month = {dec},
  
	publisher = {American Physical Society ({APS})},
  
	volume = {96},
  
	number = {11},
  
	author = {Simon Knapen and Tongyan Lin and Kathryn M. Zurek},
  
	title = {Light dark matter: Models and constraints},
  
	journal = {Physical Review D}
}

@article{Song:2021yar,
    author = "Song, Ningqiang and Nagorny, Serge and Vincent, Aaron C.",
    title = "{Pushing the frontier of WIMPy inelastic dark matter: Journey to the end of the periodic table}",
    eprint = "2104.09517",
    archivePrefix = "arXiv",
    primaryClass = "hep-ph",
    doi = "10.1103/PhysRevD.104.103032",
    journal = "Phys. Rev. D",
    volume = "104",
    number = "10",
    pages = "103032",
    year = "2021"
}

@article{M_nster_2014,
	doi = {10.1088/1475-7516/2014/05/018},
  
	url = {https://doi.org/10.1088%2F1475-7516%2F2014%2F05%2F018},
  
	year = 2014,
	month = {may},
  
	publisher = {{IOP} Publishing},
  
	volume = {2014},
  
	number = {05},
  
	pages = {018--018},
  
	author = {A Münster and M. v Sivers and G Angloher and A Bento and C Bucci and L Canonica and A Erb and F. v Feilitzsch and P Gorla and A Gütlein and D Hauff and J Jochum and H Kraus and J.-C Lanfranchi and M Laubenstein and J Loebell and Y Ortigoza and F Petricca and W Potzel and F Pröbst and J Puimedon and F Reindl and S Roth and K Rottler and C Sailer and K Schäffner and J Schieck and S Scholl and S Schönert and W Seidel and L Stodolsky and C Strandhagen and R Strauss and A Tanzke and M Uffinger and A Ulrich and I Usherov and S Wawoczny and M Willers and M Wüstrich and A Zöller},
    title = "{Radiopurity of CaWO4 crystals for direct dark matter search with CRESST and EURECA}"
    %	title = {Radiopurity of {CaWO}$\less$sub$\greater$4$\less$/sub$\greater$ crystals for direct dark matter search with {CRESST} and {EURECA}
},
  
	journal = {Journal of Cosmology and Astroparticle Physics}
}

@article{Knapen_2022,
	doi = {10.1103/physrevd.105.015014},
  
	url = {https://doi.org/10.1103%2Fphysrevd.105.015014},
  
	year = 2022,
	month = {jan},
  
	publisher = {American Physical Society ({APS})},
  
	volume = {105},
  
	number = {1},
  
	author = {Simon Knapen and Jonathan Kozaczuk and Tongyan Lin},
  
	title = {python package for dark matter scattering in dielectric targets},
  
	journal = {Physical Review D}
}

@article{Karachentsev_2012,
   title={Missing dark matter in the local universe},
   volume={67},
   ISSN={1990-3421},
   url={http://dx.doi.org/10.1134/S1990341312020010},
   DOI={10.1134/s1990341312020010},
   number={2},
   journal={Astrophysical Bulletin},
   publisher={Pleiades Publishing Ltd},
   author={Karachentsev, I. D.},
   year={2012},
   month={Apr},
   pages={123–134}
}

@article{Herrera_2021,
   title={Direct detection of non-galactic light dark matter},
   volume={820},
   ISSN={0370-2693},
   url={http://dx.doi.org/10.1016/j.physletb.2021.136551},
   DOI={10.1016/j.physletb.2021.136551},
   journal={Physics Letters B},
   publisher={Elsevier BV},
   author={Herrera, Gonzalo and Ibarra, Alejandro},
   year={2021},
   month={Sep},
   pages={136551}
}

@article{Makarov:2010di,
    author = "Makarov, Dmitry and Karachentsev, Igor",
    title = "{Galaxy groups and clouds in the Local (z $\sim$ 0.01) universe}",
    eprint = "1011.6277",
    archivePrefix = "arXiv",
    primaryClass = "astro-ph.CO",
    doi = "10.1111/j.1365-2966.2010.18071.x",
    journal = "Mon. Not. Roy. Astron. Soc.",
    volume = "412",
    pages = "2498",
    year = "2011"
}

@article{Baushev:2012dm,
    author = "Baushev, A.N.",
    title = "{Extragalactic dark matter and direct detection experiments}",
    eprint = "1208.0392",
    archivePrefix = "arXiv",
    primaryClass = "astro-ph.CO",
    doi = "10.1088/0004-637X/771/2/117",
    journal = "Astrophys. J.",
    volume = "771",
    pages = "117",
    year = "2013"
}

@article{Read:2014qva,
    author = "Read, J. I.",
    title = "{The Local Dark Matter Density}",
    eprint = "1404.1938",
    archivePrefix = "arXiv",
    primaryClass = "astro-ph.GA",
    reportNumber = "JPHYSG-100038.R1",
    doi = "10.1088/0954-3899/41/6/063101",
    journal = "J. Phys. G",
    volume = "41",
    pages = "063101",
    year = "2014"
}

@article{Green:2011bv,
    author = "Green, Anne M.",
    title = "{Astrophysical uncertainties on direct detection experiments}",
    eprint = "1112.0524",
    archivePrefix = "arXiv",
    primaryClass = "astro-ph.CO",
    doi = "10.1142/S0217732312300042",
    journal = "Mod. Phys. Lett. A",
    volume = "27",
    pages = "1230004",
    year = "2012"
}

@article{CarrilloGonzalez:2021lxm,
    author = "Carrillo Gonz\'alez, Mariana and Toro, Natalia",
    title = "{Cosmology and signals of light pseudo-Dirac dark matter}",
    eprint = "2108.13422",
    archivePrefix = "arXiv",
    primaryClass = "hep-ph",
    reportNumber = "Imperial/TP/2021/MC/03",
    doi = "10.1007/JHEP04(2022)060",
    journal = "JHEP",
    volume = "04",
    pages = "060",
    year = "2022"
}

@article{Aprile_2018,
	doi = {10.1103/physrevlett.121.111302},
  
	url = {https://doi.org/10.1103%2Fphysrevlett.121.111302},
  
	year = 2018,
	month = {sep},
  
	publisher = {American Physical Society ({APS})},
  
	volume = {121},
  
	number = {11},
  
	author = {E. Aprile and J. Aalbers and F. Agostini and M. Alfonsi and L. Althueser and F.{\hspace{0.167em}
}D. Amaro and M. Anthony and F. Arneodo and L. Baudis and B. Bauermeister and M.{\hspace{0.167em}}L. Benabderrahmane and T. Berger and P.{\hspace{0.167em}}A. Breur and A. Brown and A. Brown and E. Brown and S. Bruenner and G. Bruno and R. Budnik and C. Capelli and J.{\hspace{0.167em}}M.{\hspace{0.167em}}R. Cardoso and D. Cichon and D. Coderre and A.{\hspace{0.167em}}P. Colijn and J. Conrad and J.{\hspace{0.167em}}P. Cussonneau and M.{\hspace{0.167em}}P. Decowski and P. de Perio and P. Di Gangi and A. Di Giovanni and S. Diglio and A. Elykov and G. Eurin and J. Fei and A.{\hspace{0.167em}}D. Ferella and A. Fieguth and W. Fulgione and A. Gallo Rosso and M. Galloway and F. Gao and M. Garbini and C. Geis and L. Grandi and Z. Greene and H. Qiu and C. Hasterok and E. Hogenbirk and J. Howlett and R. Itay and F. Joerg and B. Kaminsky and S. Kazama and A. Kish and G. Koltman and H. Landsman and R.{\hspace{0.167em}}F. Lang and L. Levinson and Q. Lin and S. Lindemann and M. Lindner and F. Lombardi and J.{\hspace{0.167em}}A.{\hspace{0.167em}}M. Lopes and J. Mahlstedt and A. Manfredini and T. Marrod{\'{a}}n Undagoitia and J. Masbou and D. Masson and M. Messina and K. Micheneau and K. Miller and A. Molinario and K. Mor{\aa} and M. Murra and J. Naganoma and K. Ni and U. Oberlack and B. Pelssers and F. Piastra and J. Pienaar and V. Pizzella and G. Plante and R. Podviianiuk and N. Priel and D. Ram{\'{\i}}rez Garc{\'{\i}}a and L. Rauch and S. Reichard and C. Reuter and B. Riedel and A. Rizzo and A. Rocchetti and N. Rupp and J.{\hspace{0.167em}}M.{\hspace{0.167em}}F. dos Santos and G. Sartorelli and M. Scheibelhut and S. Schindler and J. Schreiner and D. Schulte and M. Schumann and L. Scotto Lavina and M. Selvi and P. Shagin and E. Shockley and M. Silva and H. Simgen and D. Thers and F. Toschi and G. Trinchero and C. Tunnell and N. Upole and M. Vargas and O. Wack and H. Wang and Z. Wang and Y. Wei and C. Weinheimer and C. Wittweg and J. Wulf and J. Ye and Y. Zhang and T. Zhu and},
  
	title = {Dark Matter Search Results from a One Ton-Year Exposure of {XENON}1T},
  
	journal = {Physical Review Letters}
}

@article{Angloher_2016,
	doi = {10.1140/epjc/s10052-016-3877-3},
  
	url = {https://doi.org/10.1140%2Fepjc%2Fs10052-016-3877-3},
  
	year = 2016,
	month = {jan},
  
	publisher = {Springer Science and Business Media {LLC}
},
  
	volume = {76},
  
	number = {1},
  
	author = {G. Angloher and A. Bento and C. Bucci and L. Canonica and X. Defay and A. Erb and F. von Feilitzsch and N. Ferreiro Iachellini and P. Gorla and A. Gütlein and D. Hauff and J. Jochum and M. Kiefer and H. Kluck and H. Kraus and J. C. Lanfranchi and J. Loebell and A. Münster and C. Pagliarone and F. Petricca and W. Potzel and F. Pröbst and F. Reindl and K. Schäffner and J. Schieck and S. Schönert and W. Seidel and L. Stodolsky and C. Strandhagen and R. Strauss and A. Tanzke and H. H. Trinh Thi and C. Türko{\u{g}}lu and M. Uffinger and A. Ulrich and I. Usherov and S. Wawoczny and M. Willers and M. Wüstrich and A. Zöller},
  
	title = {Results on light dark matter particles with a low-threshold {CRESST}-{II} detector},
  
	journal = {The European Physical Journal C}
}

@article{Brenner:2022qku,
    author = "Brenner, Anja and Herrera, Gonzalo and Ibarra, Alejandro and Kang, Sunghyun and Scopel, Stefano and Tomar, Gaurav",
    title = "{Complementarity of experiments in probing the non-relativistic effective theory of dark matter-nucleon interactions}",
    eprint = "2203.04210",
    archivePrefix = "arXiv",
    primaryClass = "hep-ph",
    doi = "10.1088/1475-7516/2022/06/026",
    journal = "JCAP",
    volume = "06",
    number = "06",
    pages = "026",
    year = "2022"
}

@inproceedings{Essig:2022dfa,
    author = "Essig, Rouven and Giovanetti, Graham K. and Kurinsky, Noah and McKinsey, Dan and Ramanathan, Karthik and Stifter, Kelly and Yu, Tien-Tien",
    title = "{Snowmass2021 Cosmic Frontier: The landscape of low-threshold dark matter direct detection in the next decade}",
    booktitle = "{2022 Snowmass Summer Study}",
    eprint = "2203.08297",
    archivePrefix = "arXiv",
    primaryClass = "hep-ph",
    reportNumber = "FERMILAB-CONF-22-181-PPD",
    month = "3",
    year = "2022"
}

@article{Barello_2014,
	doi = {10.1103/physrevd.90.094027},
  
	url = {https://doi.org/10.1103%2Fphysrevd.90.094027},
  
	year = 2014,
	month = {nov},
  
	publisher = {American Physical Society ({APS})},
  
	volume = {90},
  
	number = {9},
  
	author = {G. Barello and Spencer Chang and Christopher A. Newby},
  
	title = {A model independent approach to inelastic dark matter scattering},
  
	journal = {Physical Review D}
}

@article{Schwetz_2011,
	doi = {10.1088/1475-7516/2011/08/008},
  
	url = {https://doi.org/10.1088%2F1475-7516%2F2011%2F08%2F008},
  
	year = 2011,
	month = {aug},
  
	publisher = {{IOP} Publishing},
  
	volume = {2011},
  
	number = {08},
  
	pages = {008--008},
  
	author = {Thomas Schwetz and Jure Zupan},
  
	title = {Dark matter attempts for {CoGeNT} and {DAMA}
},
  
	journal = {Journal of Cosmology and Astroparticle Physics}
}

@article{Arkani_Hamed_2009,
	doi = {10.1103/physrevd.79.015014},
  
	url = {https://doi.org/10.1103%2Fphysrevd.79.015014},
  
	year = 2009,
	month = {jan},
  
	publisher = {American Physical Society ({APS})},
  
	volume = {79},
  
	number = {1},
  
	author = {Nima Arkani-Hamed and Douglas P. Finkbeiner and Tracy R. Slatyer and Neal Weiner},
  
	title = {A theory of dark matter},
  
	journal = {Physical Review D}
}

@article{Chang_2010,
	doi = {10.1103/physrevd.82.125011},
  
	url = {https://doi.org/10.1103%2Fphysrevd.82.125011},
  
	year = 2010,
	month = {dec},
  
	publisher = {American Physical Society ({APS})},
  
	volume = {82},
  
	number = {12},
  
	author = {Spencer Chang and Neal Weiner and Itay Yavin},
  
	title = {Magnetic inelastic dark matter},
  
	journal = {Physical Review D}
}

@article{Hall_1998,
	doi = {10.1016/s0370-2693(98)00196-8},
  
	url = {https://doi.org/10.1016%2Fs0370-2693%2898%2900196-8},
  
	year = 1998,
	month = {apr},
  
	publisher = {Elsevier {BV}
},
  
	volume = {424},
  
	number = {3-4},
  
	pages = {305--312},
  
	author = {Lawrence J. Hall and Takeo Moroi and Hitoshi Murayama},
  
	title = {Sneutrino cold dark matter with lepton-number violation},
  
	journal = {Physics Letters B}
}

@article{Emken_2022,
	doi = {10.1103/physrevd.105.055023},
  
	url = {https://doi.org/10.1103%2Fphysrevd.105.055023},
  
	year = 2022,
	month = {mar},
  
	publisher = {American Physical Society ({APS})},
  
	volume = {105},
  
	number = {5},
  
	author = {Timon Emken and Jonas Frerick and Saniya Heeba and Felix Kahlhoefer},
  
	title = {Electron recoils from terrestrial upscattering of inelastic dark matter},
  
	journal = {Physical Review D}
}

@article{Radick:2020qip,
    author = "Radick, Aria and Taki, Anna-Maria and Yu, Tien-Tien",
    title = "{Dependence of Dark Matter - Electron Scattering on the Galactic Dark Matter Velocity Distribution}",
    eprint = "2011.02493",
    archivePrefix = "arXiv",
    primaryClass = "hep-ph",
    doi = "10.1088/1475-7516/2021/02/004",
    journal = "JCAP",
    volume = "02",
    pages = "004",
    year = "2021"
}

@article{Kerr:1986hz,
    author = "Kerr, F. J. and Lynden-Bell, Donald",
    title = "{Review of galactic constants}",
    journal = "Mon. Not. Roy. Astron. Soc.",
    volume = "221",
    pages = "1023",
    year = "1986"
}

@article{Smith:2006ym,
    author = "Smith, Martin C. and others",
    title = "{The RAVE Survey: Constraining the Local Galactic Escape Speed}",
    eprint = "astro-ph/0611671",
    archivePrefix = "arXiv",
    doi = "10.1111/j.1365-2966.2007.11964.x",
    journal = "Mon. Not. Roy. Astron. Soc.",
    volume = "379",
    pages = "755--772",
    year = "2007"
}

@article{Piffl:2013mla,
    author = "Piffl, Til and others",
    title = "{The RAVE survey: the Galactic escape speed and the mass of the Milky Way}",
    eprint = "1309.4293",
    archivePrefix = "arXiv",
    primaryClass = "astro-ph.GA",
    doi = "10.1051/0004-6361/201322531",
    journal = "Astron. Astrophys.",
    volume = "562",
    pages = "A91",
    year = "2014"
}

@article{McCabe:2013kea,
    author = "McCabe, Christopher",
    title = "{The Earth's velocity for direct detection experiments}",
    eprint = "1312.1355",
    archivePrefix = "arXiv",
    primaryClass = "astro-ph.CO",
    reportNumber = "IPPP-13-95, DCPT-13-190",
    doi = "10.1088/1475-7516/2014/02/027",
    journal = "JCAP",
    volume = "02",
    pages = "027",
    year = "2014"
}

@article{LZ:2022ufs,
    author = "Aalbers, J. and others",
    collaboration = "LZ",
    title = "{First Dark Matter Search Results from the LUX-ZEPLIN (LZ) Experiment}",
    eprint = "2207.03764",
    archivePrefix = "arXiv",
    primaryClass = "hep-ex",
    month = "7",
    year = "2022"
}

@article{Nagata_2015_2,
	doi = {10.1103/physrevd.91.055035},
  
	url = {https://doi.org/10.1103%2Fphysrevd.91.055035},
  
	year = 2015,
	month = {mar},
  
	publisher = {American Physical Society ({APS})},
  
	volume = {91},
  
	number = {5},
  
	author = {Natsumi Nagata and Satoshi Shirai},
  
	title = {Electroweakly interacting Dirac dark matter},
  
	journal = {Physical Review D}
}

@article{Barak_2020,
	doi = {10.1103/physrevlett.125.171802},
  
	url = {https://doi.org/10.1103%2Fphysrevlett.125.171802},
  
	year = 2020,
	month = {oct},
  
	publisher = {American Physical Society ({APS})},
  
	volume = {125},
  
	number = {17},
  
	author = {Liron Barak and Itay M. Bloch and Mariano Cababie and Gustavo Cancelo and Luke Chaplinsky and Fernando Chierchie and Michael Crisler and Alex Drlica-Wagner and Rouven Essig and Juan Estrada and Erez Etzion and Guillermo Fernandez Moroni and Daniel Gift and Sravan Munagavalasa and Aviv Orly and Dario Rodrigues and Aman Singal and Miguel Sofo Haro and Leandro Stefanazzi and Javier Tiffenberg and Sho Uemura and Tomer Volansky and Tien-Tien Yu and},
  
	title = {{SENSEI}: Direct-Detection Results on sub-{GeV} Dark Matter from a New Skipper {CCD}
},
  
	journal = {Physical Review Letters}
}

@article{Catena_2020,
	doi = {10.1103/physrevresearch.2.033195},
  
	url = {https://doi.org/10.1103%2Fphysrevresearch.2.033195},
  
	year = 2020,
	month = {aug},
  
	publisher = {American Physical Society ({APS})},
  
	volume = {2},
  
	number = {3},
  
	author = {Riccardo Catena and Timon Emken and Nicola A. Spaldin and Walter Tarantino},
  
	title = {Atomic responses to general dark matter-electron interactions},
  
	journal = {Physical Review Research}
}

@article{Amole_2016,
    author = "Amole, C. and others",
    collaboration = "PICO",
    title = "{Dark matter search results from the PICO-60 CF$_3$I bubble chamber}",
    eprint = "1510.07754",
    archivePrefix = "arXiv",
    primaryClass = "hep-ex",
    reportNumber = "FERMILAB-PUB-15-449-AE-E",
    doi = "10.1103/PhysRevD.93.052014",
    journal = "Phys. Rev. D",
    volume = "93",
    number = "5",
    pages = "052014",
    year = "2016",
    url = {https://doi.org/10.1103%2Fphysrevd.93.052014},
}

@article{Biswas:2022cyh,
    author = "Biswas, Anirban and Kar, Arpan and Kim, Hyomin and Scopel, Stefano and Velasco-Sevilla, Liliana",
    title = "{Improved White Dwarves Constraints on Inelastic Dark Matter and Left-Right Symmetric Models}",
    eprint = "2206.06667",
    archivePrefix = "arXiv",
    primaryClass = "hep-ph",
    reportNumber = "CQUeST-2022-0691, KIAS P22038",
    month = "6",
    year = "2022"
}

@article{Bell_2018,
	doi = {10.1088/1475-7516/2018/09/018},
  
	url = {https://doi.org/10.1088%2F1475-7516%2F2018%2F09%2F018},
  
	year = 2018,
	month = {sep},
  
	publisher = {{IOP} Publishing},
  
	volume = {2018},
  
	number = {09},
  
	pages = {018--018},
  
	author = {Nicole F. Bell and Giorgio Busoni and Sandra Robles},
  
	title = {Heating up neutron stars with inelastic dark matter},
  
	journal = {Journal of Cosmology and Astroparticle Physics}
}

@article{Fujiwara:2022uiq,
    author = "Fujiwara, Motoko and Hamaguchi, Koichi and Nagata, Natsumi and Zheng, Jiaming",
    title = "{Capture of Electroweak Multiplet Dark Matter in Neutron Stars}",
    eprint = "2204.02238",
    archivePrefix = "arXiv",
    primaryClass = "hep-ph",
    month = "4",
    year = "2022"
}

@article{Essig:2017kqs,
    author = "Essig, Rouven and Volansky, Tomer and Yu, Tien-Tien",
    title = "{New Constraints and Prospects for sub-GeV Dark Matter Scattering off Electrons in Xenon}",
    eprint = "1703.00910",
    archivePrefix = "arXiv",
    primaryClass = "hep-ph",
    reportNumber = "CERN-TH-2017-042, YITP-SB-17-09",
    doi = "10.1103/PhysRevD.96.043017",
    journal = "Phys. Rev. D",
    volume = "96",
    number = "4",
    pages = "043017",
    year = "2017"
}

@ARTICLE{1959ApJ...130..705K,
       author = {{Kahn}, F.~D. and {Woltjer}, L.},
        title = "{Intergalactic Matter and the Galaxy.}",
      journal = "Astrophys. J.",
         year = 1959,
       volume = {130},
        pages = {705},
          doi = {10.1086/146762},
}

@BOOK{2008gady.book.....B,
       author = {{Binney}, James and {Tremaine}, Scott},
        title = "{Galactic Dynamics: Second Edition}",
         year = 2008,
       adsurl = {https://ui.adsabs.harvard.edu/abs/2008gady.book.....B}
}

@article{Cox:2007nt,
    author = "Cox, T. J. and Loeb, Abraham",
    title = "{The Collision Between The Milky Way And Andromeda}",
    eprint = "0705.1170",
    archivePrefix = "arXiv",
    primaryClass = "astro-ph",
    doi = "10.1111/j.1365-2966.2008.13048.x",
    journal = "Mon. Not. Roy. Astron. Soc.",
    volume = "386",
    pages = "461",
    year = "2008"
}

@article{Karachentsev:2002wf,
    author = "Karachentsev, I. D. and others",
    title = "{Local galaxy flows within 5 mpc}",
    eprint = "astro-ph/0211011",
    archivePrefix = "arXiv",
    doi = "10.1051/0004-6361:20021566",
    journal = "Astron. Astrophys.",
    volume = "398",
    pages = "479--492",
    year = "2003"
}

@article{OHare:2019qxc,
    author = "O'Hare, Ciaran A. J. and Evans, N. Wyn and McCabe, Christopher and Myeong, GyuChul and Belokurov, Vasily",
    title = "{Velocity substructure from Gaia and direct searches for dark matter}",
    eprint = "1909.04684",
    archivePrefix = "arXiv",
    primaryClass = "astro-ph.GA",
    doi = "10.1103/PhysRevD.101.023006",
    journal = "Phys. Rev. D",
    volume = "101",
    number = "2",
    pages = "023006",
    year = "2020"
}

@article{OHare:2018trr,
    author = "O'Hare, Ciaran A. J. and McCabe, Christopher and Evans, N. Wyn and Myeong, GyuChul and Belokurov, Vasily",
    title = "{Dark matter hurricane: Measuring the S1 stream with dark matter detectors}",
    eprint = "1807.09004",
    archivePrefix = "arXiv",
    primaryClass = "astro-ph.CO",
    doi = "10.1103/PhysRevD.98.103006",
    journal = "Phys. Rev. D",
    volume = "98",
    number = "10",
    pages = "103006",
    year = "2018"
}

@article{Bozorgnia:2019mjk,
    author = "Bozorgnia, Nassim and Fattahi, Azadeh and Frenk, Carlos S. and Cheek, Andrew and Cerdeno, David G. and G\'omez, Facundo A. and Grand, Robert J. J. and Marinacci, Federico",
    title = "{The dark matter component of the Gaia radially anisotropic substructure}",
    eprint = "1910.07536",
    archivePrefix = "arXiv",
    primaryClass = "astro-ph.GA",
    doi = "10.1088/1475-7516/2020/07/036",
    journal = "JCAP",
    volume = "07",
    pages = "036",
    year = "2020"
}

@article{Besla:2019xbx,
    author = "Besla, Gurtina and Peter, Annika and Garavito-Camargo, Nicolas",
    title = "{The highest-speed local dark matter particles come from the Large Magellanic Cloud}",
    eprint = "1909.04140",
    archivePrefix = "arXiv",
    primaryClass = "astro-ph.GA",
    doi = "10.1088/1475-7516/2019/11/013",
    journal = "JCAP",
    volume = "11",
    pages = "013",
    year = "2019"
}

@article{Smith-Orlik:2023kyl,
    author = "Smith-Orlik, Adam and others",
    title = "{The impact of the Large Magellanic Cloud on dark matter direct detection signals}",
    eprint = "2302.04281",
    archivePrefix = "arXiv",
    primaryClass = "astro-ph.GA",
    month = "2",
    year = "2023"
}

@article{Maity:2022enp,
    author = "Maity, Tarak Nath and Laha, Ranjan",
    title = "{Dark matter substructures affect dark matter-electron scattering in xenon-based direct detection experiments}",
    eprint = "2208.14471",
    archivePrefix = "arXiv",
    primaryClass = "hep-ph",
    doi = "10.1007/JHEP02(2023)200",
    journal = "JHEP",
    volume = "02",
    pages = "200",
    year = "2023"
}

@article{Freese:2003na,
    author = "Freese, Katherine and Gondolo, Paolo and Newberg, Heidi Jo and Lewis, Matthew",
    title = "{The effects of the Sagittarius dwarf tidal stream on dark matter detectors}",
    eprint = "astro-ph/0310334",
    archivePrefix = "arXiv",
    doi = "10.1103/PhysRevLett.92.111301",
    journal = "Phys. Rev. Lett.",
    volume = "92",
    pages = "111301",
    year = "2004"
}

@article{Buch:2020xyt,
    author = "Buch, Jatan and Buen-Abad, Manuel A. and Fan, Jiji and Leung, John Shing Chau",
    title = "{Dark Matter Substructure under the Electron Scattering Lamppost}",
    eprint = "2007.13750",
    archivePrefix = "arXiv",
    primaryClass = "hep-ph",
    doi = "10.1103/PhysRevD.102.083010",
    journal = "Phys. Rev. D",
    volume = "102",
    number = "8",
    pages = "083010",
    year = "2020"
}

@article{Kelso:2016qqj,
    author = "Kelso, Chris and Savage, Christopher and Valluri, Monica and Freese, Katherine and Stinson, Gregory S. and Bailin, Jeremy",
    title = "{The impact of baryons on the direct detection of dark matter}",
    eprint = "1601.04725",
    archivePrefix = "arXiv",
    primaryClass = "astro-ph.GA",
    reportNumber = "NORDITA-2015-15, CETUP2015-030",
    doi = "10.1088/1475-7516/2016/08/071",
    journal = "JCAP",
    volume = "08",
    pages = "071",
    year = "2016"
}

@article{Bozorgnia:2016ogo,
    author = "Bozorgnia, Nassim and Calore, Francesca and Schaller, Matthieu and Lovell, Mark and Bertone, Gianfranco and Frenk, Carlos S. and Crain, Robert A. and Navarro, Julio F. and Schaye, Joop and Theuns, Tom",
    title = "{Simulated Milky Way analogues: implications for dark matter direct searches}",
    eprint = "1601.04707",
    archivePrefix = "arXiv",
    primaryClass = "astro-ph.CO",
    doi = "10.1088/1475-7516/2016/05/024",
    journal = "JCAP",
    volume = "05",
    pages = "024",
    year = "2016"
}
\end{document}